\documentclass{emulateapj}
\slugcomment{Accepted for publication in ApJ}

\shorttitle {Accreting millisecond pulsar waveforms}
\shortauthors {Lamb et al.}

\begin{document}

\title {A model for the waveform behavior of accreting millisecond
pulsars:\\ Nearly aligned magnetic fields and moving emission regions}

\author {Frederick K. Lamb\altaffilmark{1,2}, Stratos
Boutloukos\altaffilmark{1,3}, Sandor Van
Wassenhove\altaffilmark{1}, Robert T.
Chamberlain\altaffilmark{1}, Ka Ho Lo\altaffilmark{1}, Alexander
Clare\altaffilmark{1}, Wenfei Yu\altaffilmark{1}, and M. Coleman
Miller\altaffilmark{3}}

\affil{{$^1$}{Center for Theoretical Astrophysics and
Department of Physics, University of Illinois at
Urbana-Champaign, 1110 West Green Street, Urbana, IL 61801-3080, USA; fkl@illinois.edu}\\
{$^2$}{Department of Astronomy, University of Illinois at
Urbana-Champaign, 1110 West Green Street, Urbana, IL 61801-3074, USA}
{$^3$}{Department of Astronomy and Maryland Astronomy Center
for Theory and Computation, University of Maryland, 
College Park, MD 20742-2421, USA}}

\begin{abstract}
\noindent  We investigate further a model of the accreting millisecond
X-ray pulsars we proposed earlier. In this model, the X-ray--emitting
regions of these pulsars are near their spin axes but move. This is to
be expected if the magnetic poles of these stars are close to their spin
axes, so that accreting gas is channeled there. As the accretion rate
and the structure of the inner disk vary, gas is channeled along
different field lines to different locations on the stellar surface,
causing the X-ray--emitting areas to move. We show that this ``nearly
aligned moving spot model'' can explain many properties of the accreting
millisecond X-ray pulsars, including their generally low oscillation
amplitudes and nearly sinusoidal waveforms; the variability of their
pulse amplitudes, shapes, and phases; the correlations in this
variability; and the similarity of the accretion- and nuclear-powered
pulse shapes and phases in some. It may also explain why
accretion-powered millisecond pulsars are difficult to detect, why some
are intermittent, and why all detected so far are transients. This model
can be tested by comparing with observations the waveform changes it
predicts, including the changes with accretion rate.
\end{abstract}


\section{Introduction}
\label{sec:intro}

Highly periodic millisecond X-ray oscillations have been detected with
high confidence in 22 accreting neutron stars in low-mass X-ray binary
systems (\mbox{LMXBs}), using the \textit {Rossi X-ray Timing Explorer}
(\textit {RXTE}) satellite (see~\citealt {lamb08a}). We refer to these
stars as accreting millisecond X-ray pulsars (\mbox{AMXPs}).
Accretion-powered millisecond oscillations have so far been detected in
10 \mbox{AMXPs}. They are always observable in seven \mbox{AMXPs}, but
are only intermittently detected in three others. Nuclear-powered
millisecond oscillations have been detected with high confidence during
thermonuclear X-ray bursts in 16 \mbox{AMXPs}. Persistent
accretion-powered millisecond oscillations have been detected in two
\mbox{AMXPs} that produce nuclear-powered millisecond oscillations;
intermittent accretion-powered millisecond oscillations have been
detected in two others.

The \mbox{AMXPs} have several important properties:

\textit {Low oscillation amplitudes}. The fractional amplitudes of the
accretion-powered oscillations of most \mbox{AMXPs} are often only
$\sim\,$1\%--2\%.\footnote{We characterize the strengths of oscillations
by their rms amplitudes, because the rms amplitude can be defined for
any waveform, is usually relatively stable, and is closely related to
the power. We convert reported semi-amplitudes of purely sinusoidal
oscillations or Fourier components to rms amplitudes by dividing by
$\sqrt2$.} Persistent accretion-powered oscillations with amplitudes
$\la\,$1\% are often detected with high confidence in IGR~J00291$+$5934
(\citealt{gall05,patr08}) and XTE~J1751$-$305 (\citealt{mark02,
patr08}). Persistent accretion-powered oscillations with amplitudes as
low as 2\% are regularly seen in XTE~J1807$-$294 (\citealt{zhan06,
chou08, patr08}), XTE~J0929$-$314 (\citealt{gall02}), and
XTE~J1814$-$338 (\citealt{chun08, patr08}). The amplitude of the
accretion-powered oscillation seen in SWIFT~1756.9$-$2508 was
$\sim\,$6\% (\citealt{krim07}). The intermittent accretion-powered
oscillations detected in SAX~J1748.9$-$2021 (\citealt{gavr07, alta08,
patr08}), HETE~J1900.1$-$2455 (\citealt{gall07}), and Aql~X-1
(\citealt{case08}) have amplitudes $\sim\,$0.5\%--3\%.

\textit {Nearly sinusoidal waveforms}. The waveforms (light curves) of
the accretion-powered oscillations of most \mbox{AMXPs} are nearly
sinusoidal (see \citealt{wijn06} and the references in the preceding
paragraph). The amplitude of the first harmonic (fundamental) component
is usually $\ga\,$10 times the amplitude of the second harmonic (first
overtone) component, although the ratio can be as small as $\sim\,$3.5,
as is sometimes the case in XTE~J1807$-$294 (\citealt{zhan06}), or even
$\ga\,$1, as is sometimes the case in SAX~J1808.4$-$3658 (see, e.g.,
\citealt{hart08}).

\textit {Highly variable oscillation amplitudes}. The fractional
amplitudes of the accretion-powered oscillations of most \mbox{AMXPs}
vary in time by factors ranging from $\sim\,$2 to $\sim\,$10. Observed
fractional amplitudes vary from 0.7\% to 1.7\% in SAX~J1748.9$-$2021,
from 0.7\% to 3.7\% in XTE~J1751$-$305, from 3\% to 7\% in
XTE~J0929$-$314, from 1\% to 9\% in IGR~J00291$+$5934, from 2\% to 11\%
in XTE~J1814$-$338, from 1\% to 14\% in XTE~J1808$-$338, and from 2\% to
19\% in XTE~J1807$-$294 (see the references above).

\textit {Highly variable pulse phases}. The phases of accretion-powered
pulses have been seen to vary rapidly by as much as $\sim\,$0.3 cycles
in several \mbox{AMXPs}, including SAX~J1808.4$-$3658 \citep{morg03,
hart08} and XTE~J1807$-$294 \citep{mark04}. Wild changes in the apparent
pulse frequency have been observed with \textit {both} signs at the
\textit{same} accretion rate in XTE~J1807$-$294 (see \citealt{mark04}).
If interpreted as caused by changes in the stellar spin rate, these
phase variations would be more than a factor of 10 larger than expected
for the largest accretion torques and smallest inertial moments thought
possible for these systems (see \citealt{ghos79b, latt01}).

\textit {Undetected accretion-powered oscillations}. More than 80
accreting neutron stars in \mbox{LMXBs} are known (\citealt{chak05,
liu07}), but accretion-powered millisecond X-ray oscillations have so
far been detected in only 10 of them. Accretion-powered oscillations
have not yet been detected even in 13 \mbox{AMXPs} that produce periodic
nuclear-powered millisecond oscillations, indicating that they have
millisecond spin periods (\citealt{lamb08a}); eight of these also
produce kilohertz quasi-periodic oscillations (\mbox{QPOs}) with
frequency separations that indicate that they not only have millisecond
spin periods but also have dynamically important magnetic fields
(\citealt{bout08a}).

\textit {Intermittent accretion-powered oscillations}. Accretion-powered
millisecond X-ray pulsations have been detected only occasionally in
SAX~J1748.9$-$2021 (\citealt{gavr07, alta08, patr08}),
HETE~J1900.1$-$2455 (\citealt{gall07}), and Aql~X-1 (\citealt{case08}).
When oscillations are not detected, the upper limits are typically
$\la0.5$\%.

\textit{Correlated pulse arrival times and amplitudes}. The phase
residuals of the accretion-powered pulses of several \mbox{AMXPs} appear
to be anti-correlated with their fractional amplitudes, at least over
some of the amplitude ranges observed. \mbox{AMXPs} that show this type
of behavior include XTE~J1807$-$294 and XTE~J1814$-$338
(\citealt{patr08}).

\textit{Similar accretion- and nuclear-powered pulses}. The shapes and
phases of the nuclear-powered X-ray pulses of the \mbox{AMXPs}
SAX~J1808.4$-$3658 (\citealt{chak03}) and XTE~J1814$-$338
(\citealt{stro03}) are very similar to the  shapes and phases of their
accretion-powered X-ray pulses.

\textit {Concentration in transient systems of \mbox{AMXPs} with
accretion-powered oscillations}. The \mbox{AMXPs} in which
accretion-powered oscillations have been detected tend to be found in
binary systems that have outbursts lasting about a month (but see
\citealt{gall08}) separated by quiescent intervals lasting years
(\citealt{chak05, rigg08}). The accretion rates of these neutron stars
are very low.

In this paper we investigate further the ``nearly aligned moving spot''
model of \mbox{AMXP} X-ray emission that we proposed previously
\citep{lamb06,lamb07,lamb08b}. This model has three main features:

\begin{enumerate}
\item  The strongest poles of the magnetic fields of neutron stars with
millisecond spin periods are located near---and sometimes very
near---the stellar spin axis. This behavior is expected for several
magnetic field evolution mechanisms.

\item  The star's magnetic field channels accreting gas close to its
spin axis, creating X-ray emitting areas there and depositing nuclear
fuel there.\footnote{\cite{muno02} considered a single bright spot near
the spin axis as well as a uniformly bright hemisphere and antipodal
spots near the spin equator as possible reasons for the nearly
sinusoidal waveforms of some X-ray burst oscillations, but did not
consider accretion-powered oscillations or other consequences of
emission from near the spin axis.}

\item  The X-ray emitting areas on the stellar surface move, as changes
in the accretion rate and the structure of the inner disk cause
accreting gas to be channeled along different field lines to different
locations on the stellar surface. (The magnetic field of the neutron
star is fixed in the stellar crust on the timescales relevant to the
phenomena considered here.)
\end{enumerate}

These features provide the basic ingredients needed to understand the
\mbox{AMXP} properties discussed above. This is the subject of the
sections that follow. As a guide to these sections, we summarize our
results here.

\begin{enumerate}
\item  Emission from near the spin axis naturally produces weak
modulation, regardless of the viewing direction. The reason is that
uniform emission from a spot centered on the spin axis is axisymmetric
about the spin axis and therefore produces no modulation. Emission from
a spot close to the spin axis has only a small asymmetry and therefore
produces only weak modulation.

\item  Emission from near the spin axis also naturally produces a nearly
sinusoidal waveform, because the asymmetry of the emission is weak and
broad.

\item  If the emitting area is close to the spin axis, even a small
movement in latitude can change the oscillation amplitude by a
substantial factor.

\item  If the emitting area is close to the spin axis, movement in the
longitudinal direction by a small distance can change the phase of the
oscillation by a large amount.

Changes in the latitude and longitude of the emitting area are expected
on timescales at least as short as the $\sim\,$0.1~ms dynamical time at
the stellar surface and as long as the $\sim\,$10~d timescale of the
variations observed in the mass accretion rate.

\item  If the emitting area is very close to the spin axis and remains
there, the oscillation amplitude may be so low that it is undetectable.
The effects of rapid changes in the position of the emitting
area---possibly in combination with other effects, such as reduction of
the modulation fraction by scattering in circumstellar gas---may also
play a role in reducing the detectability of accretion-powered
oscillations in neutron stars with millisecond spin periods. These
effects may explain the fact that accretion-powered X-ray oscillations
have not yet been detected in many accreting neutron stars that are
thought to have millisecond spin periods and dynamically important
magnetic fields.

\item  If the emitting area is very close to the spin axis, a small
change in the accretion flow can suddenly channel gas farther from the
spin axis, causing the emitting area to move away from the axis. This
can make a previously undetectable oscillation become detectable.
Temporary motion of the emitting area away from the spin axis may
explain the intermittent accretion-powered oscillations of some
\mbox{AMXPs} (\citealt{lamb09}).

\item  If the pulse amplitude and phase variations observed in
\mbox{AMXPs} are caused by motion of the emitting area, they should be
correlated. In particular, the pulse phase should be much more scattered
when the pulse amplitude is very low. The reason is that changes in the
longitudinal position of the emitting area by a given distance produce
much larger phase changes when the emitting area is very close to the
spin axis, which is also when the oscillation amplitude is very low.

The observational consequences discussed so far depend only on
features~(2) and~(3) of the model, i.e., that the accretion-powered
X-ray emission of \mbox{AMXPs} comes from areas near their spin axes and
that these areas move significantly on timescales of hours to days.

\item  The picture of \mbox{AMXP} X-ray emission outlined here suggests
that the shapes and phases of the nuclear- and accretion-powered pulses
are similar to one another in some \mbox{AMXPs} because the nuclear- and
accretion powered X-ray emission comes from approximately the same area
on the stellar surface. The reason for this is that in some cases, the
mechanism that concentrates the magnetic flux of the accreting neutron
star near its spin axis, as it is spun up, will naturally produce
magnetic fields strong enough to confine accreting nuclear fuel near the
magnetic poles at least partially, even though the dipole component of
the magnetic field is weak.

\item  The picture of neutron star magnetic field evolution and
\mbox{AMXP} X-ray emission outlined here also suggests a possible
explanation for why the \mbox{AMXPs} in which accretion-powered
oscillations have been detected are in transient systems. If most
neutron stars in \mbox{LMXBs} were spun up by accretion from a low spin
rate to a high spin rate, their magnetic poles were forced very close to
their spin axes, making accretion-powered oscillations difficult or
impossible to detect. However, those stars that are now in compact
transient systems now experience infrequent episodes of mass accretion
and the accretion rate is very low. By now they have been spun down from
their maximum spin rates, a process that could force their magnetic
poles away from their spin axes enough to produce detectable
accretion-powered oscillations.
\end{enumerate}

These last two observational consequences depend on feature~(1) of the
model, i.e., on how the magnetic fields of neutron stars evolve as they
are spun up and down by accretion and electromagnetic torques.

In the remainder of this paper we discuss in detail the features of the
model and its observational implications. In Section~\ref {sec:model}, we
outline our approach, discussing our modeling of X-ray emission from the
stellar surface, our computational and the code verification methods,
and the pulse profile representation we use.

We present our results in Sections~\ref {sec:amplitudes} and~\ref
{sec:variations}. These results are based on our computations of several
hundred million waveforms for different emitting regions, beaming
patterns, stellar models, and viewing directions. In Section~\ref
{sec:amplitudes}, we consider the shape and amplitude of X-ray pulses as
a function of the size and inclination of the emitting areas, the
compactness of the star, and the stellar spin rate. In Section~\ref
{sec:variations}, we consider the changes in the pulse amplitude and
phase produced by various motions of the emitting regions on the stellar
surface and explore the origins of correlated changes in the pulse
amplitude and phase and the effects of rapid movement of the emitting
areas. We also discuss why oscillations have not yet been detected in
many accreting neutron stars in LMXBs and why accretion-powered
oscillations are detected only intermittently in some \mbox{AMXPs}.

In Section~\ref {sec:discussion}, we summarize the results of our model
calculations. We also discuss how the magnetic poles of most
\mbox{AMXPs} can be forced close to their spin axes, how such mechanisms
may explain why the \mbox{AMXPs} that produce accretion-powered
millisecond oscillations are transient pulsars, the consistency of the
model with the observed properties of rotation-powered millisecond
pulsars, and possible observational tests of the model discussed here.

Further results of our investigation of the present model will be
presented elsewhere (S. Boutloukos et al., in preparation).

\section{X-ray Waveform Modeling}
\label{sec:model}

\subsection {Modeling the X-ray Emission}
\label {sec:modeling-emission}

In the radiating spot model of \mbox{AMXP} X-ray emission, the waveform
seen by a distant observer depends on the sizes, shapes, and positions
of the emitting regions on the stellar surface; the beaming pattern of
the radiation; the compactness, radius, and spin rate of the star; and
the direction from which the star is observed. The properties of the
X-ray emitting regions are determined by the strength and geometry of
the star's magnetic field, the locations where plasma from the accretion
disk enters the magnetosphere, the extent to which the accreting plasma
becomes threaded and channeled by the stellar magnetic field, and the
resulting plasma flow pattern onto the stellar surface.

In principle, accreting plasma can reach the stellar surface in two
basic ways: (1)~by becoming threaded by the stellar magnetic field and
then guided along stellar field lines to the vicinity of a stellar
magnetic pole (\citealt{lamb73, bask75, elsn76, ghos77, ghos79a,
ghos79b}) or (2)~by penetrating between lines of the stellar magnetic
field via the magnetic version of the Rayleigh--Taylor instability
(\citealt{lamb75a, lamb75b, elsn76, elsn77, aron76, lamb77}) and then
spiraling inward to the stellar surface.

If a centered dipole component is the strongest component of the star's
magnetic field, plasma in the accretion disk that becomes threaded and
then channeled to the vicinity of a magnetic pole is expected to impact
the star in a partial or complete annulus around the pole, producing a
crescent- or ring-shaped emitting area near the pole. If the axis of the
dipole field is significantly tilted relative to the spin axis and the
spin axis is aligned with the axis of the accretion disk, a
crescent-shaped emitting region is expected (see, e.g., \citealt{bask75,
bask76, ghos77, daum96, mill96, mill98b, roma03}). If instead the dipole
axis is very close to the spin axis, as in the model of \mbox{AMXP}
X-ray emission proposed here, the emitting region may completely
encircle the spin axis (see, e.g., \citealt{ghos79a, ghos79b, roma03}).

The north and south magnetic poles of some \mbox{AMXPs} may be very
close to the same spin pole, producing a very off-center dipole moment
orthogonal to the spin axis (see \citealt{chen93}; \citealt{chen98}; and
Section~\ref{sec:pole-movement}). If so, accreting matter will be
channeled close to the spin axis, but may be channeled preferentially
toward one magnetic pole, producing an emitting region with
approximately one-fold symmetry about the spin axis, or about equally
toward both poles, producing an emitting region with approximately
two-fold symmetry about the spin axis. In the first case the first
harmonic of the spin frequency is likely be the dominant harmonic in the
X-ray waveform whereas in the second case the second harmonic is likely
to dominate. Which case occurs will depend on the accretion flow through
the inner disk. In either case, the X-ray emission will come from close
to the spin axis.

The neutron stars that are \mbox{AMXPs} may well have even more
complicated magnetic fields, with significant quadrupole and octopole
components. Higher multipole components are likely to play a more
important role in the \mbox{AMXPs} than in the classic strong-field
accretion-powered pulsars, because the magnetic fields of the
\mbox{AMXPs} are much weaker. As a result, accreting plasma can
penetrate closer to the stellar surface, where the higher multipole
moments of the star's magnetic field have a greater influence on the
channeling of accreting plasma (\citealt{elsn76}). In this case, plasma
will still tend to be channeled toward regions on the surface where the
magnetic field is strongest and will tend to impact the surface in rings
or annuli, but the emission pattern may be spatially complex and vary
rapidly in time (\citealt{long08}).

Disk plasma that penetrates between lines of the stellar magnetic field
will continue to drift inward as it loses its angular momentum, probably
predominantly via its interaction with the star's magnetic field
(\citealt{lamb01}). Cold plasma will remain in the disk plane and impact
the star in an annulus where the disk plane intersects the stellar
surface (\citealt{mill98b, lamb01}). If some of the accreting plasma
were to become hot, the forces exerted on it by the stellar magnetic
field would tend to drive it toward the star's magnetic equator
(\citealt{mich77}), causing it to impact the stellar surface in an
annulus around the star's magnetic equator. However, emission and
inverse Compton scattering of radiation is likely to keep the accreting
plasma cold (\citealt{elsn84}), so that it remains in the disk plane as
it drifts inward. Plasma that penetrates to the stellar surface via the
magnetic Rayleigh--Taylor instability is likely to impact the stellar
surface in rapidly fluctuating, irregular patterns (see \citealt{roma06,
roma08}).

Whether accreting plasma reaches the stellar surface predominantly via
channeled flow along field lines or via unstable flow between field
lines depends on the accretion rate and the spin frequency of the star
(see \citealt{lamb89, roma08, kulk08}). Under some conditions, plasma
may accrete in both ways simultaneously (see \citealt{mill98b, roma08,
kulk08}).

The sizes, shapes, and locations of the emitting areas on the surface of
an accreting magnetic neutron star and the properties of the emission
from these areas are expected to change on timescales at least as short
as the $\sim\,$1~ms dynamical timescale near the star. This expectation
is supported by recent simulations of accretion onto weakly magnetic
neutron stars (see \citealt{roma03, roma04, roma06, long08, roma08,
kulk08}). However, changes in \mbox{AMXP} X-ray fluxes can be measured
accurately using current instruments only by combining
$\sim\,$100--1,000~s of data and hence only variations in waveforms on
timescales longer than this can be measured directly. Consequently, the
emitting areas and beaming patterns that are relevant for comparisons
with current observations of waveforms are the averages of the actual
areas and beaming patterns over these relatively long times. The
emitting areas and beaming patterns that we use in our computations
should therefore be interpreted as averages of the actual areas and
beaming patterns over these times.

We have computed the X-ray waveforms produced by emitting regions with
various sizes, shapes, and positions, for several different X-ray
beaming patterns and a range of stellar masses, compactnesses, and spin
rates. We find that in many cases these waveforms can be approximated by
the waveforms generated by a circular, uniformly emitting spot located
at the centroid of the emitting region. The main reason for this is that
an observer sees half the star's surface at a time (or more, when
gravitational light deflection is included), which diminishes the
influence of the size and detailed shape of the emitting region on the
waveform. Consequently, we focus here on the waveforms produced by
uniformly emitting circular spots. We will discuss the waveforms
produced by emitting areas with other shapes, such as rings or
crescents, in a subsequent paper (S. Boutloukos et al., in preparation).

In addition to studying the X-ray waveforms produced by emitting areas
with fixed sizes, shapes, positions, and radiation-beaming patterns, we
are also interested in the \textit {changes} in waveforms produced by
changes in the these properties of the emitting areas. The changes we
investigate should be understood as occurring on the timescales
$\ga\,$100~s that can be studied using current instruments. It is not
yet possible to compute from first principles the accretion flows and
X-ray emission of \mbox{AMXPs} on these timescales, so simplified models
must be used. (The simulations referred to earlier follow the accretion
flow for a few dozen spin periods or dynamical times, intervals that are
orders of magnitude shorter than the intervals that are relevant).

In the following sections, we consider radiation from a single spot, from
two antipodal spots, and from two spots in the same rotational
hemisphere, near the star's spin axis. Although many uncertainties
remain, recent magnetohydrodynamic simulations of accretion onto weakly
magnetic neutron stars have found that gas impacts 1\%--20\% of the
stellar surface \citep{roma04}, equivalent to the areas of circular
spots with angular radii of 10$\arcdeg$--53$\arcdeg$. These radii are
consistent with analytical estimates of the sizes of the emission
regions of accreting neutron stars with weak magnetic fields \citep
{mill98b, psal99}. Consequently, we focus on spot sizes in this range.

An observer may see radiation from a single spot either because the
accretion flow pattern strongly favors one pole of a dipolar stellar
magnetic field over the other, or because the observer's view of one
pole is blocked by the inner disk or by accreting plasma in the star's
magnetosphere (see \citealt{mccr76, bask76}). An observer may see
radiation from two antipodal spots if emission from both magnetic poles
is visible. Finally, an observer may see radiation from two spots near
the same rotation pole if neutron vortex motion drives both of the
star's dipolar magnetic poles toward the same rotation pole (see
\citealt{chen98}).

To make it easier for the reader to compare cases, we usually report
results for our ``reference'' star, which is a $1.4\,M_\odot$ star with
a radius of $5M$ in units where $G=c=1$ (10.3~km for $M=1.4\,M_\odot$),
spinning at 400~Hz as measured at infinity, but we also discuss other
stellar models. For the same reason, we usually consider spots with
angular radii of 25$\arcdeg$. This is not an important limitation,
because the observed waveform depends only weakly on the size of the
emitting spots, as discussed in Section~\ref {sec:spot-size}. We
describe how the results change if the spot is larger or smaller.

\subsection {Computing X-ray Waveforms}
\label {sec:computing-waveforms}

The X-ray waveforms calculated here assume that radiation propagating
from emitting areas on the stellar surface reaches the observer without
interacting with any intervening matter. The bolometric X-ray waveforms
that would be seen by a distant observer were calculated using the
Schwarzschild plus Doppler (S+D) approximation introduced by
\citet{mill98a}. The S+D approximation treats exactly the special
relativistic Doppler effects (such as aberrations and energy shifts)
associated with the rotational motion of the stellar surface, but treats
the star as spherical and uses the Schwarzschild spacetime to compute
the general relativistic redshift, trace the propagation of light from
the stellar surface to the observer, and calculate light travel-time
effects. For the stars considered here, and indeed for any stars that do
not both rotate rapidly and have very low compactness, the effects of
stellar oblateness and frame dragging are minimal and are negligible
compared to uncertainties in the X-ray emission (see \citealt{cade07}).

We describe the emission from the stellar surface using coordinates
centered on the star. When considering emission from a single spot, we
denote the angle between its centroid and the star's spin axis by $i_s$
and its azimuth in the stellar coordinate system by $\phi_s$. When
considering emission from two spots, we somewhat arbitrarily identify
one as the primary spot and the other as the secondary spot. We denote
the inclination and azimuth of the centroid of the primary spot by
$i_{s1}$ and $\phi_{s1}$ and the inclination and azimuth of the centroid
of the secondary spot by $i_{s2}$ and $\phi_{s2}$. We denote the
inclination of the observer relative to the stellar spin axis by $i$.

In computing the waveforms seen by distant observers, we use as our
global coordinate system Schwarzschild coordinates $(r, \theta, \varphi,
t)$ centered on the star with $\theta=0$ aligned with the star's spin
axis and $\varphi=0$ in the plane containing the spin axis and the
observer. We choose the zero of the Schwarzschild time coordinate $t$ so
that a light pulse that propagates radially from a point on the stellar
surface immediately below the observer (i.e., at $\theta=i$ and
$\varphi=0$) arrives at the observer at $t=0$.

We carried out many calculations to test and verify the computer code
used to obtain the results we report here. We determined that the code
was giving sufficiently accurate results by varying the spatial and
angular resolutions used. For most of the cases considered in this
paper, the emitting spots were sampled by a grid of 250 points in
latitude and 250 points in longitude, the radiation-beaming pattern was
specified at 10$^4$ angles, and the flux seen by a distant observer was
computed at 10$^4$ equally spaced values of the star's rotational phase.
In some cases, finer grids were used.

We verified the code used here by comparing its results with analytical
and numerical results for several test cases:

\begin{enumerate}
\item  We tested our code's representation of emitting areas and
ray tracing in flat space by comparing the results given by our code
with exact analytical results for the absolute flux seen by an observer
directly above uniform, isotropically emitting circular spots of various
sizes. The numerical results agreed with the analytical results.

\item  We tested our code's computation of special relativistic Doppler
boosts, aberrations, and propagation time effects in several ways. We
compared the results given by our code with exact analytical results for
the waveforms produced by emission in (a)~a pencil beam normal to the
surface and (b)~a thin fan beam tangent to the surface of a rapidly
rotating star. We also compared the results given by our code with
analytical results for the waveforms produced by a small spot on the
surface of a slowly rotating star in flat space emitting
(a)~isotropically and (b)~in a beaming pattern representing Comptonized
emission (see \citealt{pout03}). The numerical results agreed with the
analytical results.

\item  We tested our code's computation of the general relativistic
redshift and light deflection for nonrotating stars by (a)~comparing the
deflection of a fan beam tangent to the stellar surface given by our
code for a variety of stellar compactnesses with the analytical
expressions for the light deflection given by \citet{pech83} and
\citet{page95}; (b)~comparing the absolute flux given by our code for an
observer directly above isotropically emitting uniform circular spots of
various sizes with independent semi-analytical results for these cases;
(c)~comparing the symmetries of the waveform and the phase of the
waveform maximum given by our code with exact analytical results for
these quantities; and (d)~comparing the shape of the waveforms given by
our code with the shapes reported by \citet{pech83} and \citet{stro92}.
Our numerical results agreed satisfactorily with the comparison results
in all cases.\footnote{The waveforms reported by \citet{pech83} for two
antipodal spots are slightly inaccurate, as shown by the following two
tests. The waveform seen by an observer in the star's rotation equator
viewing two identical antipodal spots in the rotation equator should be
the same at 180$\arcdeg$ as at 0$\arcdeg$, but this is not the case for
their waveform for this geometry (see their Figure~7). More generally,
the flux from a uniform, isotropically emitting, circular spot on a
spherical, nonrotating star should depend only on the angle between the
radius through the center of the spot and the radius to the observer.
This is not quite true for the waveforms reported by \citet{pech83}. The
waveforms given by our code pass these tests (for details, see S.
Boutloukos et al., in preparation).}

\item  We tested our code's computation of the waveforms produced by
emission from slowly rotating stars in general relativity by comparing
the rms oscillation amplitudes it gives with the amplitudes given by the
approximate analytical formulae of \citet{viir04}. Where the results of
Viironen \& Poutanen are expected to be accurate, the two sets of rms
amplitudes agreed to better than 1\%; in many cases the agreement was
much better. We also compared the waveform given by our code for an
isotropically emitting spot inclined $45\arcdeg$ from the spin axis of a
$1.4\,M_\odot$ star with a Schwarzschild coordinate radius of $5M$
spinning at 600~Hz seen by an observer at an inclination of $45\arcdeg$
with the waveform reported by \citet{pout06} for this case; the two
waveforms agreed to better than 1\%.
\end{enumerate}

Further details of these tests and comparisons will be given in a
subsequent paper (S. Boutloukos et al., in preparation).

\subsection {Constructing Pulse Profiles}
\label {sec:constructing-profiles}

The X-ray flux seen by a given observer will evolve continuously in time
as the star rotates and the emission from the stellar surface changes,
generating the observed waveform $W(t)$. As noted in Section~\ref
{sec:modeling-emission}, the accretion flow from the inner disk to the
stellar surface is expected to vary on timescales at least as short as
the $\sim\,$1~ms dynamical timescale near the neutron star, which will
cause the sizes, shapes, and positions of the emitting regions, and
therefore the observed waveform, to vary on these timescales.

The sensitivity of current instruments is too low to measure the
waveform of an \mbox{AMXP} on timescales as short as 1~ms. However,
nearly periodic waveforms with periods this short can be partially
characterized by folding segments of flux data centered at a sequence of
clock times $t_i$ (see, e.g., \citealt{hart08, patr08}). If the data are
folded with a period $P_{f}$ that is chosen to agree as closely as
possible with the local, approximate repetition period $P(t_i)$ of the
waveform, one can construct a time sequence of pulse profiles
$W_P(\phi,t_i)$; here $\phi$ is the pulse phase over one cycle.

The pulse profiles $W_P(\phi,t_i)$ constructed by folding flux data are
averages of the actual pulse profiles over the time interval required to
construct a stable profile, which can be hundreds or even thousands of
seconds, 10$^5$--10$^6$ times longer than the $\sim\,$1~ms dynamical
timescale near the neutron star. The folded pulse profiles are therefore
likely to vary more slowly and have less detail than the X-ray waveform,
a point to which we will return in Section~\ref {sec:variations}.

The waveforms of \mbox{AMXPs} can be modeled even on the dynamical
timescale near the stellar surface, but such waveforms would contain
much more information than can be studied using current observations.
Consequently, we focus here on modeling folded pulse profiles. We define
a computed pulse profile as the waveform seen by a given observer when a
star with a constant emission pattern makes one rotation. It is
customary and useful to describe pulse profiles by the amplitudes and
phases of their Fourier components. We describe our computed profiles
using the representation
\begin{equation}
W_P(t)
= A_{0} + \sum_{k=1}^{\infty} A_k \cos[2\pi k (\nu_s t - \phi_k)] \;.
\label{eqn:fourier-sum}
\end{equation}
Here $t$ is the time measured at infinity, $A_0$ is the average of the
flux over one rotation period, $A_k$ and $\phi_k$ are the amplitude and
phase of the $k$th harmonic component of the pulse, and $\nu_s$ is the
stellar rotation frequency. For harmonic $k$, the range of unique phases
is 0 to $1/k$. We sometimes also refer to the phase $\phi_p$ of the peak
of the pulse profile, defined in a similar way. With these definitions,
shifting the arrival time of a given pulse by an amount $\delta t$
shifts the phases of its Fourier components and the phase of the pulse
peak by $\nu_s\,\delta t$ cycles.

We define the arrival time $t_k$ of the $k$th harmonic component of the
pulse profile to be zero if the first maximum of the sinusoid associated
with this component reaches the observer at the same time as would a
light pulse coming from a point on the stellar surface immediately below
the observer (see Section~\ref{sec:computing-waveforms}). The arrival
time $t_k$ is affected by the sizes, shapes, and locations of the
emitting areas, the beaming pattern of the emitted radiation, and the
aberration and Doppler shifts produced by the spin of the star and
usually is not zero. We define the arrival phase of the $k$th harmonic
component by $\phi_k = \nu_s\, t_k$. The arrival time $t_p$ and arrival
phase $\phi_p$ of the peak of the pulse profile are defined similarly.
We refer to $t_k$ and $t_p$ as time residuals and to $\phi_k$ and
$\phi_p$ as phase residuals. An increase in a time or phase residual
indicates that the harmonic component or the peak is arriving later.

\section{Oscillation amplitudes}
\label{sec:amplitudes}

As discussed in Section~\ref {sec:intro}, the fractional amplitudes of
the accretion-powered oscillations of most \mbox{AMXPs} are typically
\mbox{$\sim\,$1\%}--2\%, but the amplitudes of several \mbox{AMXPs} are
sometimes as large as \mbox{$\sim\,$10\%}--20\%. A successful model of
the accretion-powered oscillations of the \mbox{AMXPs} should therefore
be able to explain oscillation amplitudes as low as $\sim\,$1\%--2\%
without requiring a special viewing angle or stellar structure and
should also be able to explain the higher amplitudes sometimes seen.

As was also discussed in Section~\ref {sec:intro}, the accretion-powered
oscillations of the \mbox{AMXPs} are often nearly sinusoidal. The second
harmonic of the fundamental oscillation frequency has been detected in
seven of the ten known \mbox{AMXPs}, but it is typically $\ga\,$10 times
weaker than the fundamental, although in a few cases it is not this weak
and in one case, SAX~J1808.4$-$3658, it is sometimes stronger than the
fundamental. Upper limits \mbox{$\sim\,$0.1\%}--0.2\% have been placed
on the amplitude of any second harmonic in three \mbox{AMXPs}.

Upper limits \mbox{$\sim\,$0.5\%}--1\% have been placed on the
fractional amplitudes of accretion-powered millisecond oscillations in
several \mbox{AMXPs} that produce nuclear-powered millisecond
oscillations, for frequencies close to the frequencies of the
nuclear-powered oscillations. A model that can explain these
nondetections as well as the observed properties of the detected
accretion-powered oscillations would be very attractive.

In this section, we show that emitting regions on or near the stellar
surface can produce oscillation amplitudes as low as $\sim\,$1\%--2\%
for a substantial range of viewing directions only if they are located
within a few degrees of the stellar spin axis. Regions near the spin
axis also naturally produce nearly sinusoidal pulse profiles. If they
are very close to the spin axis and remain there, the amplitudes of the
oscillations they produce can be $\sim\,$0.5\% or less. Thus, emission
from very close to the spin axis, in combination with other effects,
such as rapid phase and amplitude fluctuations and suppression of X-ray
modulation by scattering of photons in circumstellar
gas~\citep[see][]{lamb85, mill00}, may explain the nondetection of
accretion-powered oscillations in accreting neutron stars in which
nuclear-powered oscillations have been detected.

We show further that although the fractional oscillation amplitudes
produced by large regions tend to be smaller, this is a weak effect.
Unless almost the entire surface of the star is uniformly emitting,
amplitudes as small as those observed are possible even for large spots
only if they are located near the stellar spin axis. We also show that
although the fractional amplitudes produced by very compact neutron
stars tend to be smaller than the amplitudes produced by less compact
stars, this effect is too weak to explain the small amplitudes of the
\mbox{AMXPs}. Furthermore, it cannot explain why the fractional
amplitudes of several \mbox{AMXPs} are $\sim\,$1\%--2\% at some times
but $\sim\,$15\%--25\% a few hours or days later, since the stellar
compactness cannot change on such short timescales. Finally, we note
that the fractional amplitudes produced by spinning neutron stars depend
only weakly on their spin rates.

Measured oscillation amplitudes are likely to be smaller than those
shown in the figures in this section, which are for stable spots fixed
on the stellar surface. The reason is that the pulse shape is expected
to vary on timescales shorter than the time needed to construct a pulse
waveform. Such rapid pulse shape variations will appear as increased
background noise, reducing the apparent amplitude of the
oscillations~(see \citealt{lamb85}). As discussed below, this effect
tends to be more important when the emitting area is near the spin axis,
and it will therefore tend to reduce further the apparent amplitude of
the oscillations produced by spots at small inclinations.

In Section~\ref{sec:variations}, we show that the variations of the
oscillation amplitudes seen in all \mbox{AMXPs} and the larger
fractional amplitudes $\sim\,$15\%--20\% occasionally seen in several of
them can be explained by modest increases in the inclinations of their
emitting regions, if these regions are close to the star's spin axis.

\begin{figure*}[t]
\includegraphics[height=.185\textheight]{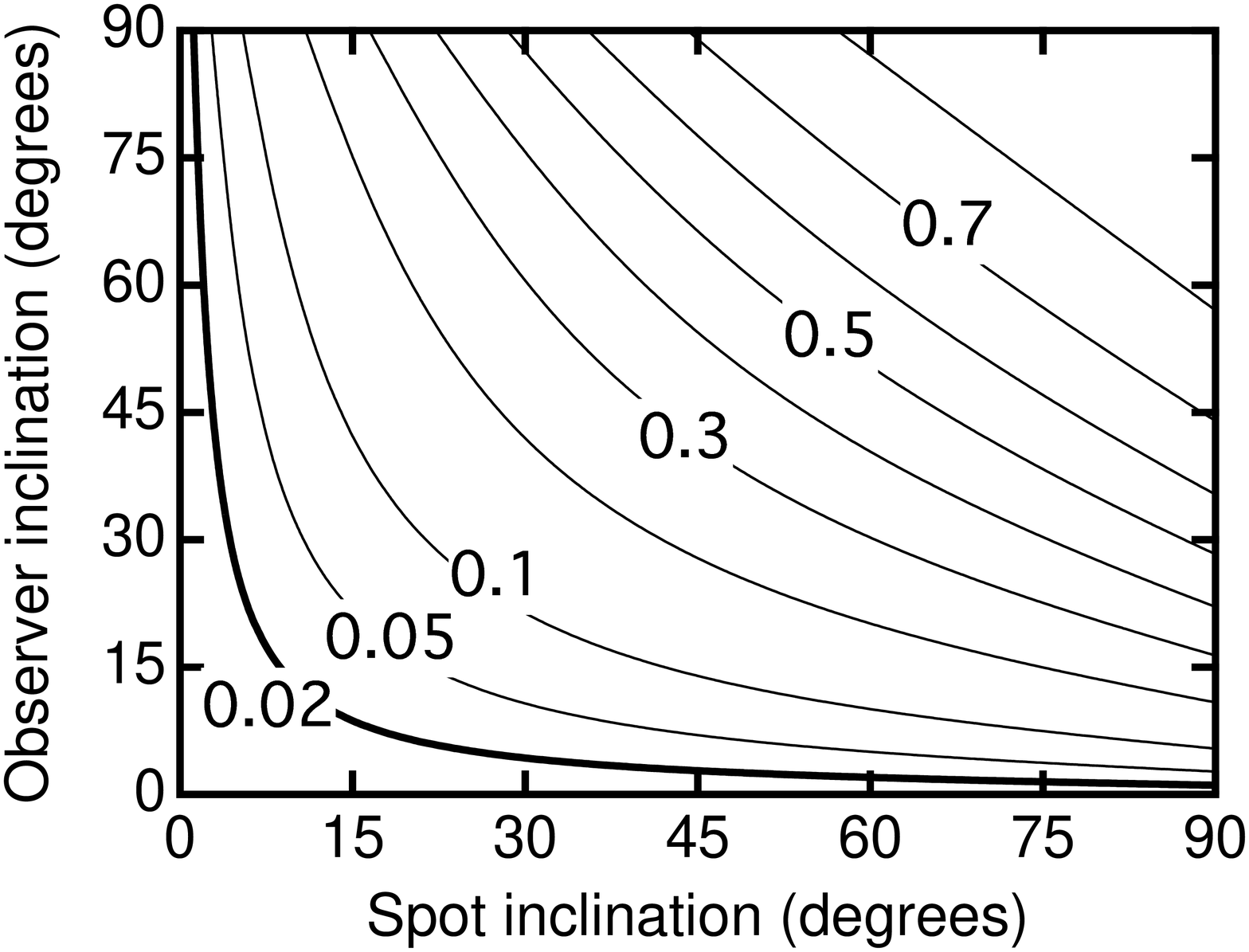}
\includegraphics[height=.185\textheight]{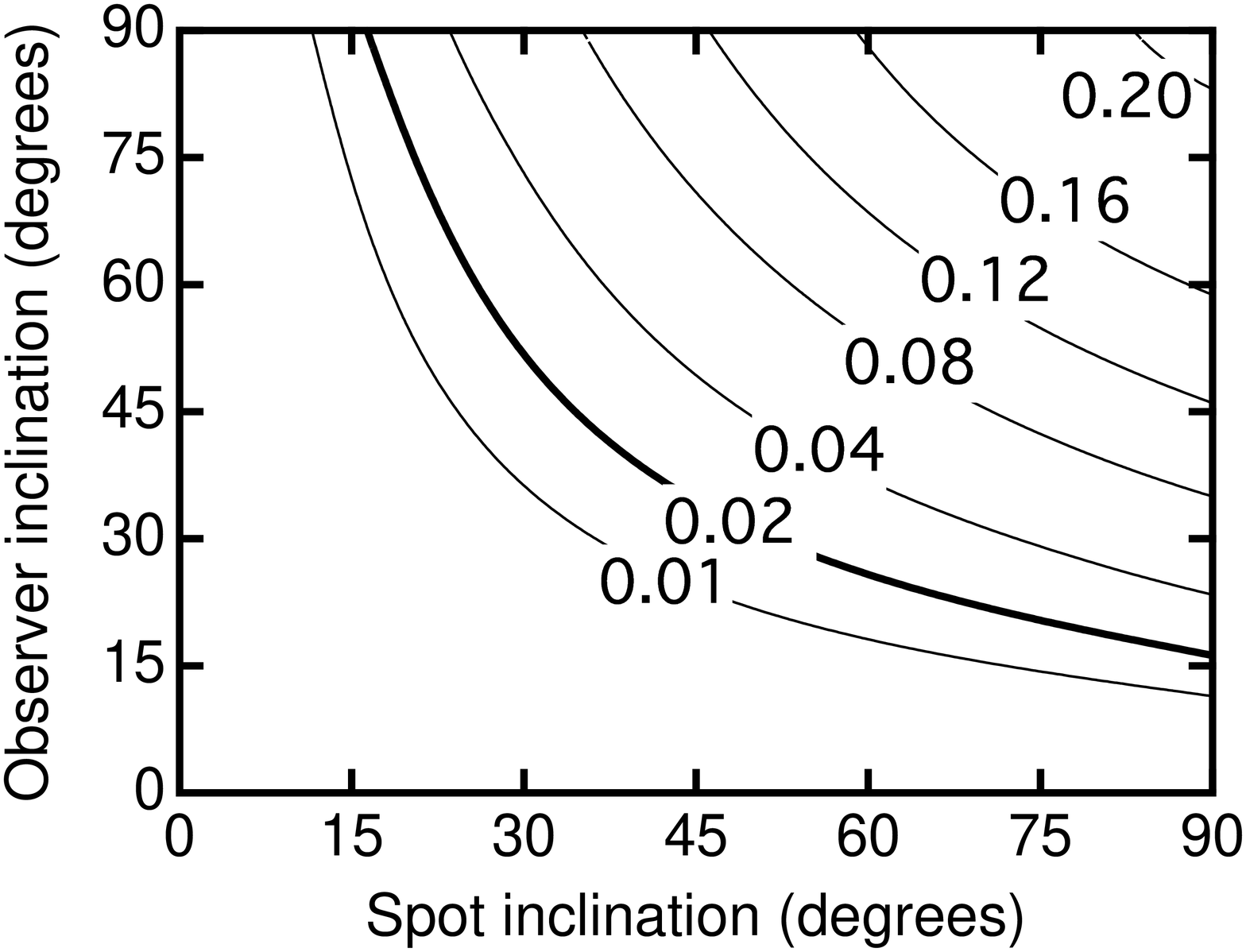}
\includegraphics[height=.185\textheight]{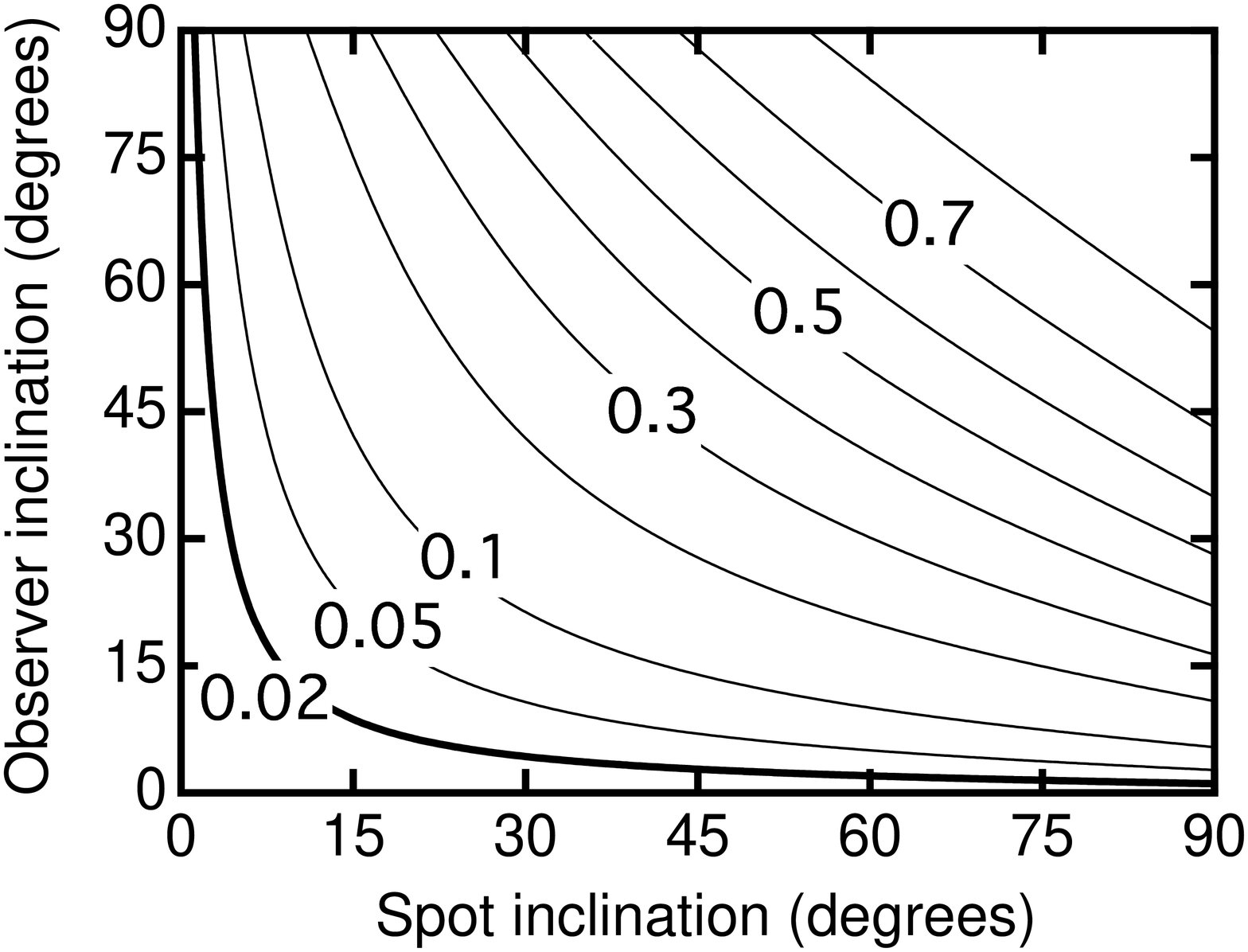}\\
\includegraphics[height=.185\textheight]{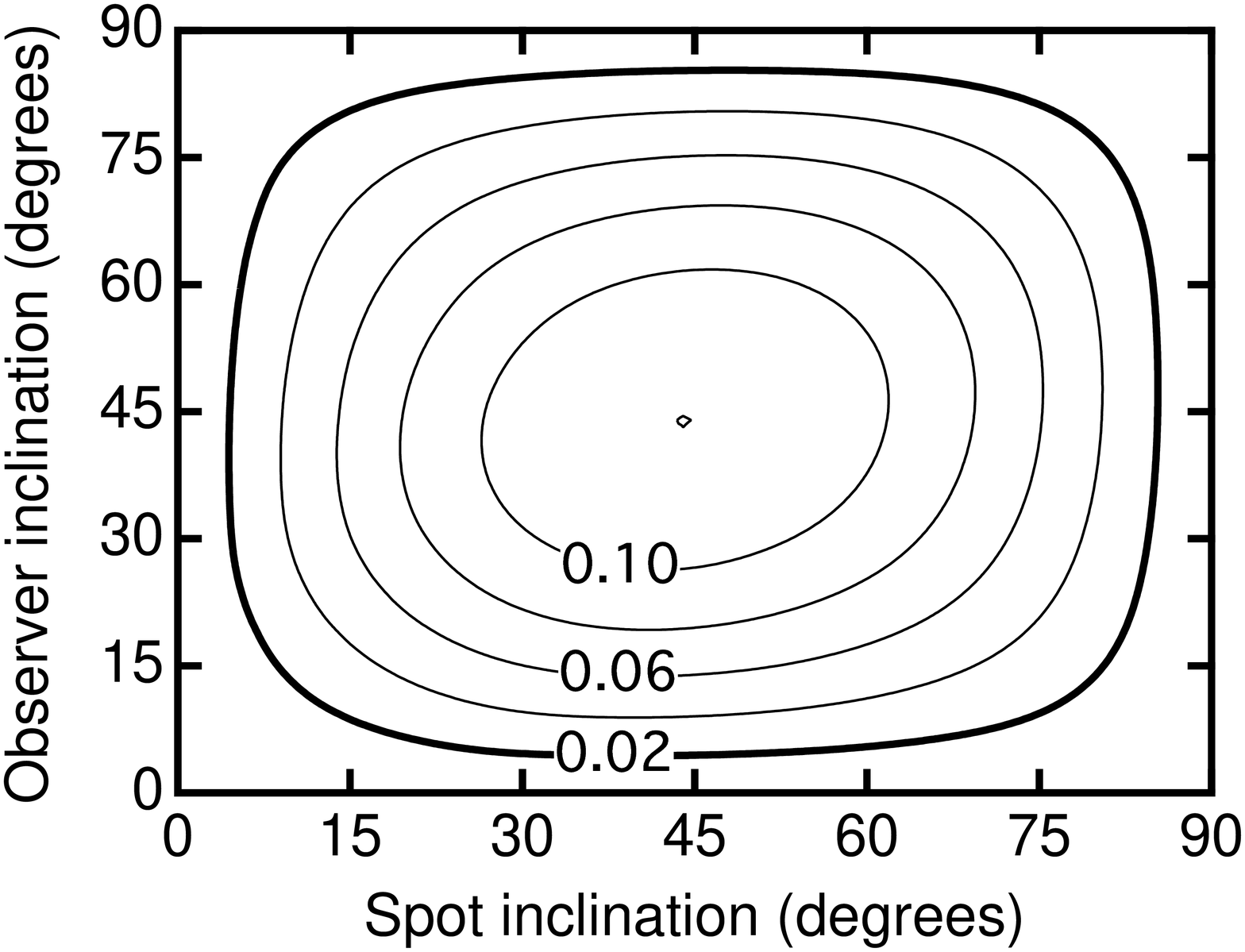}
\includegraphics[height=.185\textheight]{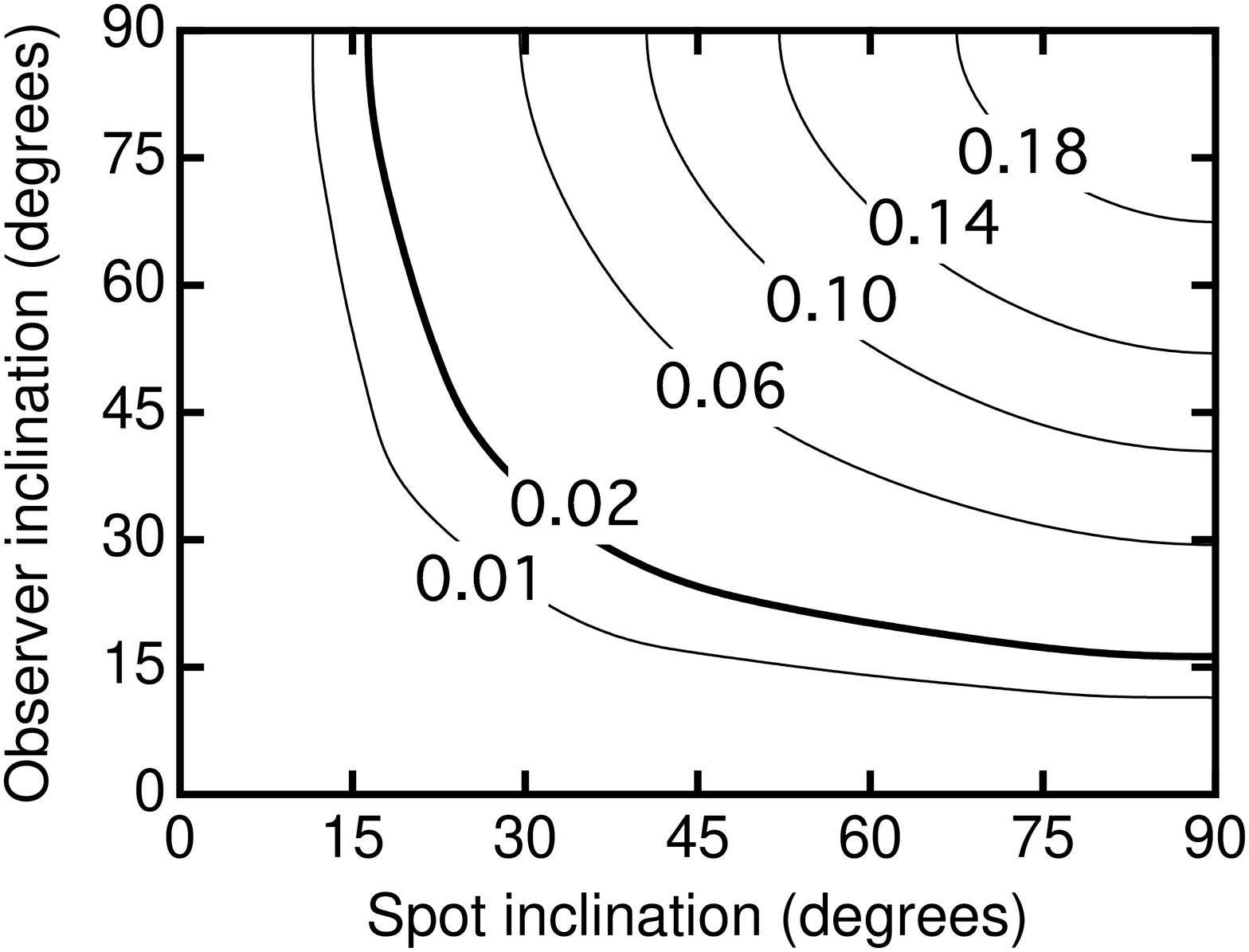}
\includegraphics[height=.185\textheight]{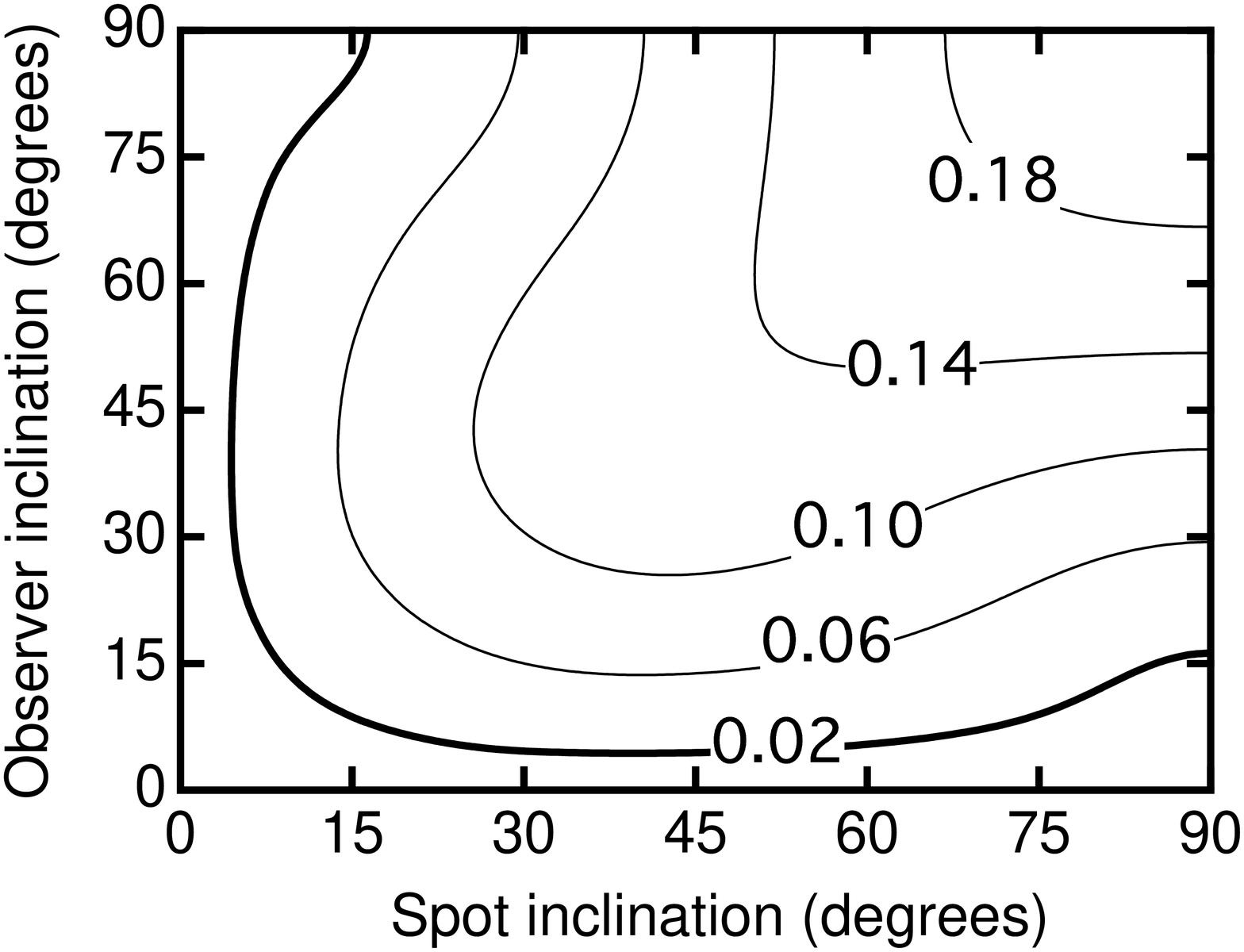}
\caption{
Top panels (left to right): contour plots of the fractional rms
amplitudes of the first harmonic (fundamental) and second harmonic
(first overtone) components of the pulse profiles and the total
fractional rms amplitude produced by a single stable spot, as a function
of the observer's inclination and the inclination of the primary spot
relative to the spin axis. Bottom panels (left to right): corresponding
contour plots for the pulse profiles produced by two stable antipodal
spots. All plots assume spot radii of 25$\arcdeg$ and a $1.4 M_\odot$
star with a radius of $5M$ spinning at $400\,$Hz. The heavier lines
highlight the contours where the fractional flux variation is 2\%. This
figure shows that emitting regions on or near the stellar surface can
produce oscillation amplitudes as low as $\sim\,$1\%--2\% for a
substantial range of viewing directions only if they are located within
a few degrees of the stellar spin axis.
}
\label{fig:amplitude-contours}
\end{figure*}

\subsection {Dependence on Spot Inclination}
\label{sec:spot-inclination}

The precise distance emitting regions can be from the spin axis and
still produce oscillation amplitudes as low as the  $\sim\,$1\%--2\%
amplitudes often observed in the \mbox{AMXPs} depends on the beaming
pattern of the emission. The angular pattern of the radiation flux
produced by Thomson scattering is strongly peaked in the direction
normal to the stellar surface, whereas the flux patterns produced by
fan-shaped radiation beams like those proposed by \citet{pout03} to
describe Comptonized emission from near the stellar surface are peaked
much less (if at all) in the direction normal to the stellar surface.
The angular pattern of the radiation flux produced by isotropic emission
is intermediate between these two cases, being peaked normal to the
surface less than the pattern produced by Thomson scattering but more
than the pattern produced by fan-beam emission. We first discuss the
waveforms produced by isotropic emission and then the waveforms produced
by other beaming patterns.

Figure~\ref{fig:amplitude-contours} shows the fractional rms amplitudes
of the first harmonic (fundamental) and second harmonic (first overtone)
components of the pulse profile and the total fractional rms amplitude
produced by isotropic emission from a single stable spot and from two
stable antipodal spots, as a function of the observer's inclination and
the inclination of the primary spot relative to the spin axis. The
curves are for spots with angular radii of $25\arcdeg$ on our reference
star (a $1.4M_\odot$ star with a radius of $5M$ spinning at $400\,$Hz).
As expected, all amplitudes are zero when the star is viewed along its
spin axis ($i=0\arcdeg$) or the emitting spots are centered on the
rotation pole ($i_s=0\arcdeg$ or $i_{s1}=0\arcdeg$ and
$i_{s2}=180\arcdeg$). In addition, the amplitude of the first harmonic
variation produced by two stable antipodal spots vanishes if either the
observer or the spots are in the rotation equator ($i_s=90\arcdeg$ or
$i_{s1}=i_{s2}=90\arcdeg$), whereas the amplitude of the second harmonic
is highest for these geometries.

The top panels in Figure~\ref{fig:amplitude-contours} show that if the
observer sees radiation only from a single spot, the total fractional
amplitude can be as low as $\sim\,$1\%--2\% for a range of viewing
angles only if the spot is within a few degrees of the spin axis. These
panels also show that for radiation from a single spot, the second
harmonic is $\sim\,$5--10 times smaller that the first harmonic for most
observer and spot inclinations.

The bottom panels in Figure~\ref{fig:amplitude-contours} show that if
the observer sees radiation from two antipodal spots, the total
fractional amplitude can be as low as $\sim\,$2\% for a substantial
range of viewing angles only if the spots are within about 5$\arcdeg$ of
the spin axis. These panels also show that for radiation from two
antipodal spots, the second harmonic is smaller than the first harmonic
for observer or spot inclinations less than about 60$\arcdeg$. The
amplitude of the second harmonic is $\la\,$2\% for all observers only if
the spots are within 15$\arcdeg$ of the spin axis.

Two emitting spots centered near the same rotation pole are expected if
the inward motion of neutron vortices in the stellar core drives both of
the star's magnetic poles toward the same rotation pole (see
Section~\ref{sec:pole-movement}). The fractional modulation produced by
two spots with this geometry is typically somewhat larger than the
modulation produced by two antipodal spots if the observer and spot
inclinations are both large, but is typically smaller than the
modulation produced by two antipodal spots if the observer inclination
or the spot inclination is small. Even so, the total fractional
amplitude can be as low as $\sim\,$1\%--2\% for a range of viewing
angles only if the spots are within a few degrees of the stellar spin
axis. For example, for our reference star the fractional modulation
produced by two spots at the same inclination but separated by
160$\arcdeg$ in longitude is larger than the modulation produced by two
antipodal spots if $i\ga70\arcdeg$ and $i_{s1}=i_{s2} \ga 70\arcdeg$ but
is smaller if $i\la60\arcdeg$ and $i_{s1}=i_{s2} \la 60\arcdeg$. The
total fractional amplitude can appear as low as $\sim\,$1\%--2\% over
half the sky ($i\ge60\arcdeg$) only if the two spots are within
$\sim\,$10$\arcdeg$ of the spin axis.

The intensity distribution produced by a Thomson scattering atmosphere
is described by a linear combination of Hopf functions (see
\citealt{chan60}, Section~68) and peaks in the direction normal to the
stellar surface. The radiation flux from such an atmosphere is given by
the product of this intensity distribution and the projected area of the
surface, which is proportional to the cosine of the angle to the normal.
Consequently, the radiation flux from a Thomson scattering surface is
strongly peaked in the direction normal to the surface. For this reason,
emission from a Thomson scattering region typically produces a higher
modulation fraction than would isotropic emission from the same region.
Hence if Thomson scattering is important, the emitting regions must be
closer to the star's spin axis than if the emission is isotropic.

Fan-shaped intensity distributions like those proposed by \citet{pout03}
to describe Comptonized emission from near the stellar surface can
partially compensate for the decrease of the projected area as the angle
from the normal to the surface increases, producing a radiation flux
distribution that is more isotropic than that produced by an isotropic
intensity distribution. For this reason, fan-shaped emission from a
single spot can produce modulation at the first harmonic of the spin
frequency that is smaller than would be produced by isotropic emission
from the same spot. A spot producing this emission pattern must still be
close to the spin axis in order for the amplitude of the first harmonic
to be $\la2$\%. Fan-shaped emission from a single spot tends to increase
the amplitude of the second harmonic relative to the first. Fan-like
emission from two antipodal spots produces waveforms similar to those
produced by isotropic emission, but the modulation fraction can be up to
a factor of 2 larger.

In SAX~J1808.4$-$3658, the second harmonic is usually weak but is
occasionally as strong as or even stronger than the first harmonic (see
\citealt{hart08}). The strengthening of the second harmonic could be
caused by development of a two-fold asymmetry of one spot, by more
fan-beam emission from a single spot, or by increased visibility a
second, antipodal spot.

\begin{figure*}[t]
\includegraphics[height=.28\textheight]{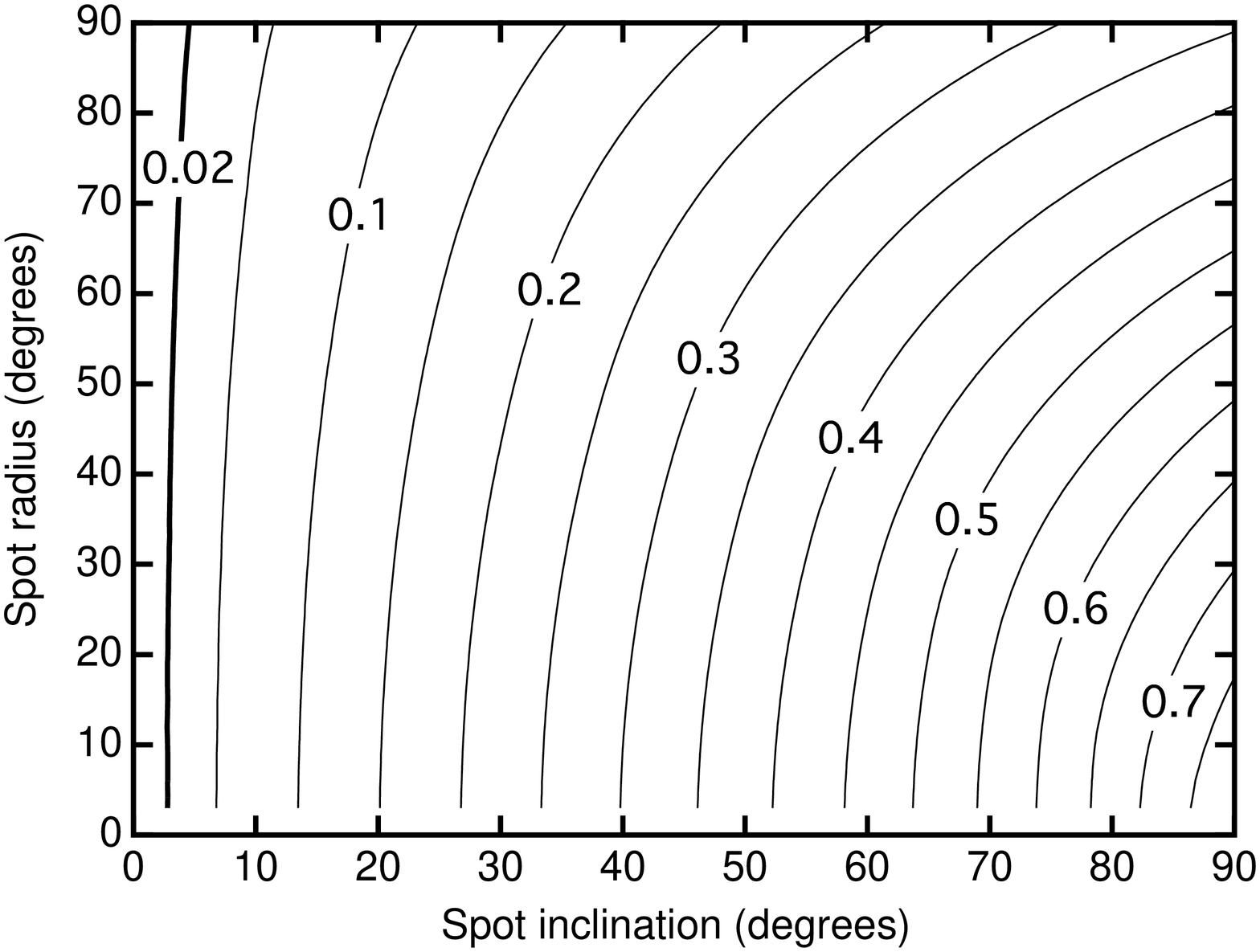}
\includegraphics[height=.28\textheight]{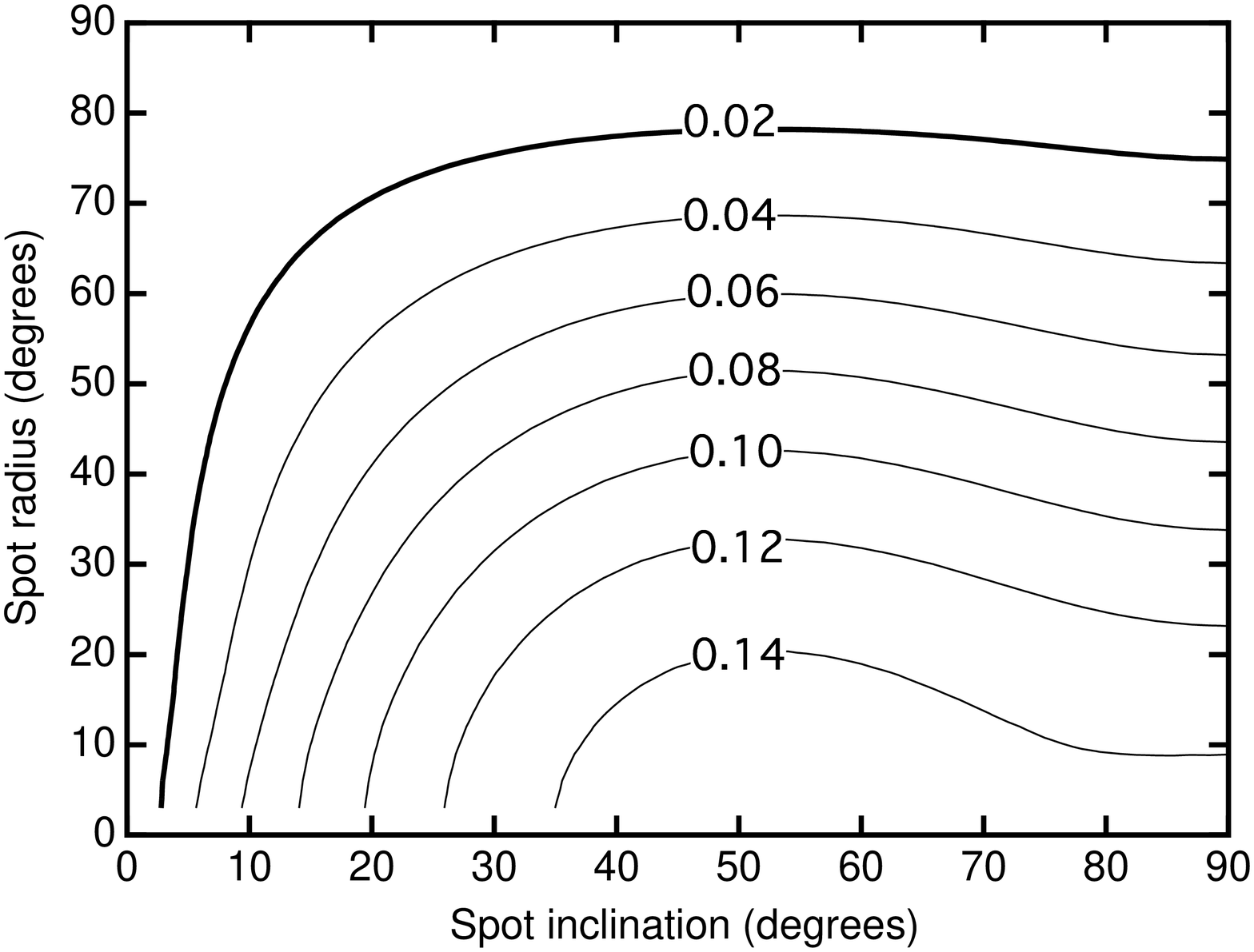}
\caption{
Contour plots of the total fractional rms amplitude produced by
isotropic emission from a single stable spot (left-hand panel) and from
two stable antipodal spots (right-hand panel), as a function of the spot
radius and the inclination of the primary spot relative to the spin
axis, for a $1.4 M_\odot$ star with a radius of $5M$ spinning at
$400\,$Hz observed at an inclination of 45$\arcdeg$. The heavier lines
highlight the contours where the total flux modulation is 2\%. This
figure demonstrates that even very large spots must be centered close to
the stellar spin axis in order to explain the low oscillation amplitudes
observed in the \mbox{AMXPs}, unless almost the entire stellar surface
is uniformly emitting.
}
\label{fig:amplitudes-spot-size}
\end{figure*}

Our computations show that modulation fractions are below typical
current detection limits if the emitting regions are within 1$\arcdeg$
of the spin axis and remain there. As discussed in Section~\ref
{sec:variations}, rapid fluctuations in the amplitude and phase of the
oscillations would not be directly detectable in time-averaged
observations, but would raise the background noise level and further
reduce the detectability of the oscillations. These effects, possibly in
combination with others, such as reduction of the modulation fraction by
scattering in circumstellar gas~\citep{lamb85, mill00}, may explain the
nondetection of accretion-powered oscillations in accreting neutron
stars in which nuclear-powered oscillations have been detected.

\subsection {Dependence on Spot Size}
\label{sec:spot-size}

Larger emitting spots tend to produce lower oscillation amplitudes than
smaller spots, but even very large spots must be centered close to the
stellar spin axis in order to explain the low oscillation amplitudes
observed in the \mbox{AMXPs}, unless almost the entire stellar surface
is uniformly emitting. This is illustrated in
Figure~\ref{fig:amplitudes-spot-size}, which shows the total fractional
amplitude produced by isotropic emission from a single spot and from two
antipodal spots, as a function of the radius of the spots and the
inclination of the primary spot relative to the spin axis. All the
curves in this figure are for our reference star, viewed at an
inclination of 45$\arcdeg$.

The left-hand panel of Figure~\ref{fig:amplitudes-spot-size} shows that
even a single spot that covers half of the star (a spot with a radius of
90$\arcdeg$) can produce a total fractional modulation as small as 2\%
only if its inclination is $\la\,$5$\arcdeg$. The modulation fraction
depends only weakly on the size of the spot, especially for
small-to-moderate spot inclinations. For example, the amplitude of the
oscillation produced by a single spot inclined 45$\arcdeg$ from the spin
axis decreases by only $\sim\,$10\% as the spot radius increases from
$5\arcdeg$ to $45\arcdeg$. Even for a spot inclined 90$\arcdeg$ from the
spin axis, the fractional flux modulation depends only weakly on the
size of the spot for spot radii $\la\,$50$\arcdeg$, decreasing from 70\%
to 35\% as the spot radius increases from 30$\arcdeg$ to 90$\arcdeg$.
The reason for this insensitivity is that the angular width of the
stellar radiation pattern produced even by spots with radii
$\sim\,$50$\arcdeg$ is governed mostly by the beaming pattern of the
radiation from the stellar surface and the magnitude of the
gravitational light deflection, not by the radius of the spot. A single
spot at an inclination of 70$\arcdeg$ can produce a fractional
modulation $\la\,$2\% only if the entire surface of the star is
uniformly emitting except for a dark area of radius 20$\arcdeg$ on the
opposite side of the star from the center of the spot (i.e., if the spot
radius is $\ga160\arcdeg$).

The right-hand panel of Figure~\ref{fig:amplitudes-spot-size} shows that
two uniformly emitting antipodal spots of a given radius produce a
smaller fractional modulation than a single spot of the same radius,
because the antipodal spots cover twice as much of the stellar surface.
Even so, the antipodal spots must be very large in order to produce
fractional modulations as small as those observed. For example, two
antipodal spots on our reference star can produce a fractional
modulation as low as $\sim\,$2\% for most observing directions only if
they have radii $\ga75\arcdeg$, which means that all of the stellar
surface is uniformly emitting except for a band around the star with a
total width $\la30\arcdeg$. The fractional modulation for most observing
directions doubles to $\sim\,$4\% if the spot radius decreases from
75$\arcdeg$ to 65$\arcdeg$.

These results show that the low fractional amplitudes often observed in
the \mbox{AMXPs} cannot be explained by large emitting regions, unless
the emitting regions are uniform and cover almost the entire stellar
surface, which is not expected. Another difficulty with attributing the
low fractional amplitudes often observed to large emitting areas is that
a substantial number of \mbox{AMXPs} that are observed to have
fractional amplitudes $\sim\,$1\%--2\% at some times are observed to
have much larger fractional amplitudes $\sim\,$15\%--20\% at other
times. Hence attributing their small amplitudes at some times to large
emitting areas would require their emitting areas to shrink by a large
factor at other times, which is not expected.

\subsection {Dependence on Stellar Compactness}
\label{sec:compactness}

\begin{figure*}[t]
\includegraphics[height=.28\textheight]{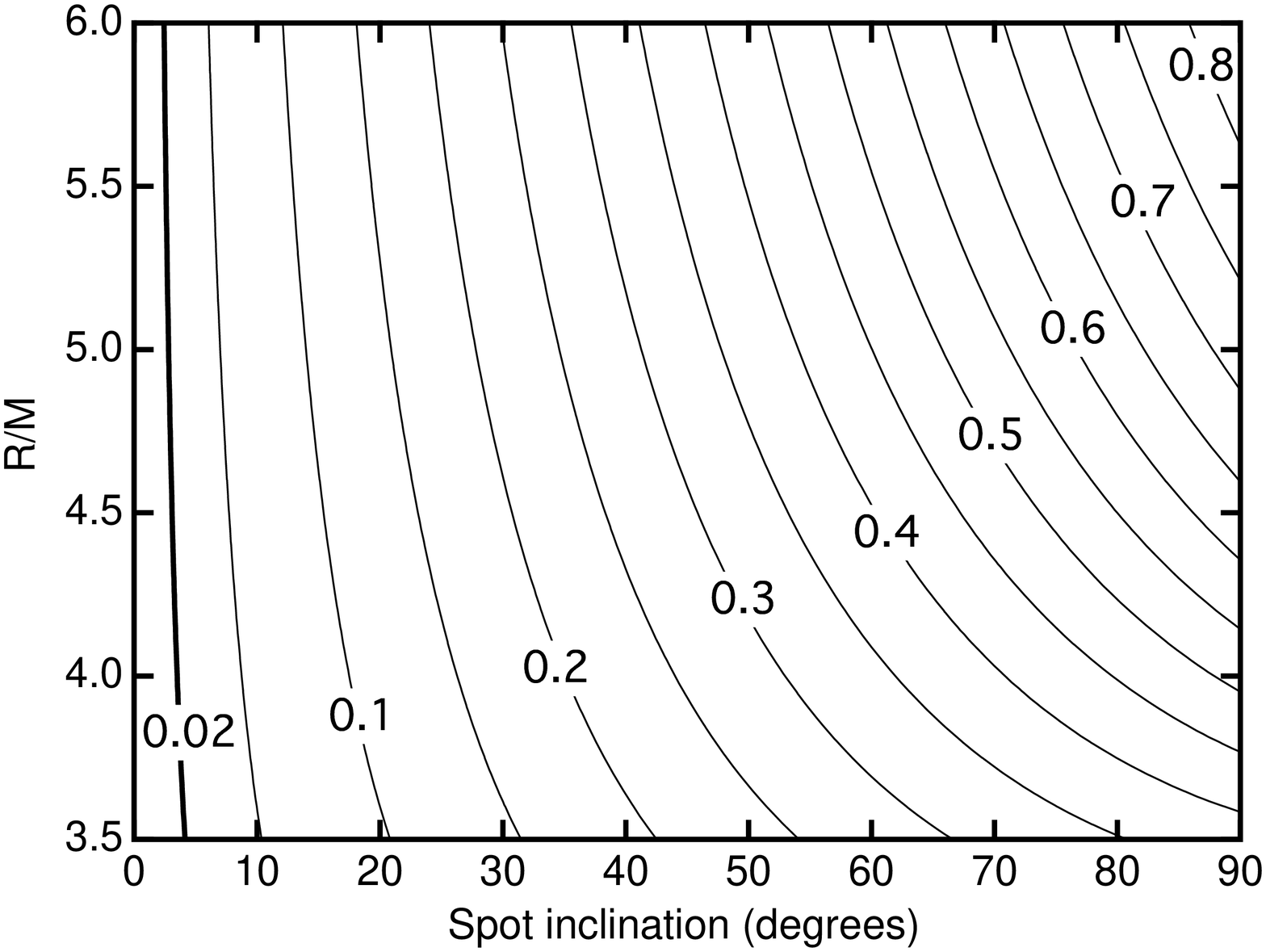}
\includegraphics[height=.28\textheight]{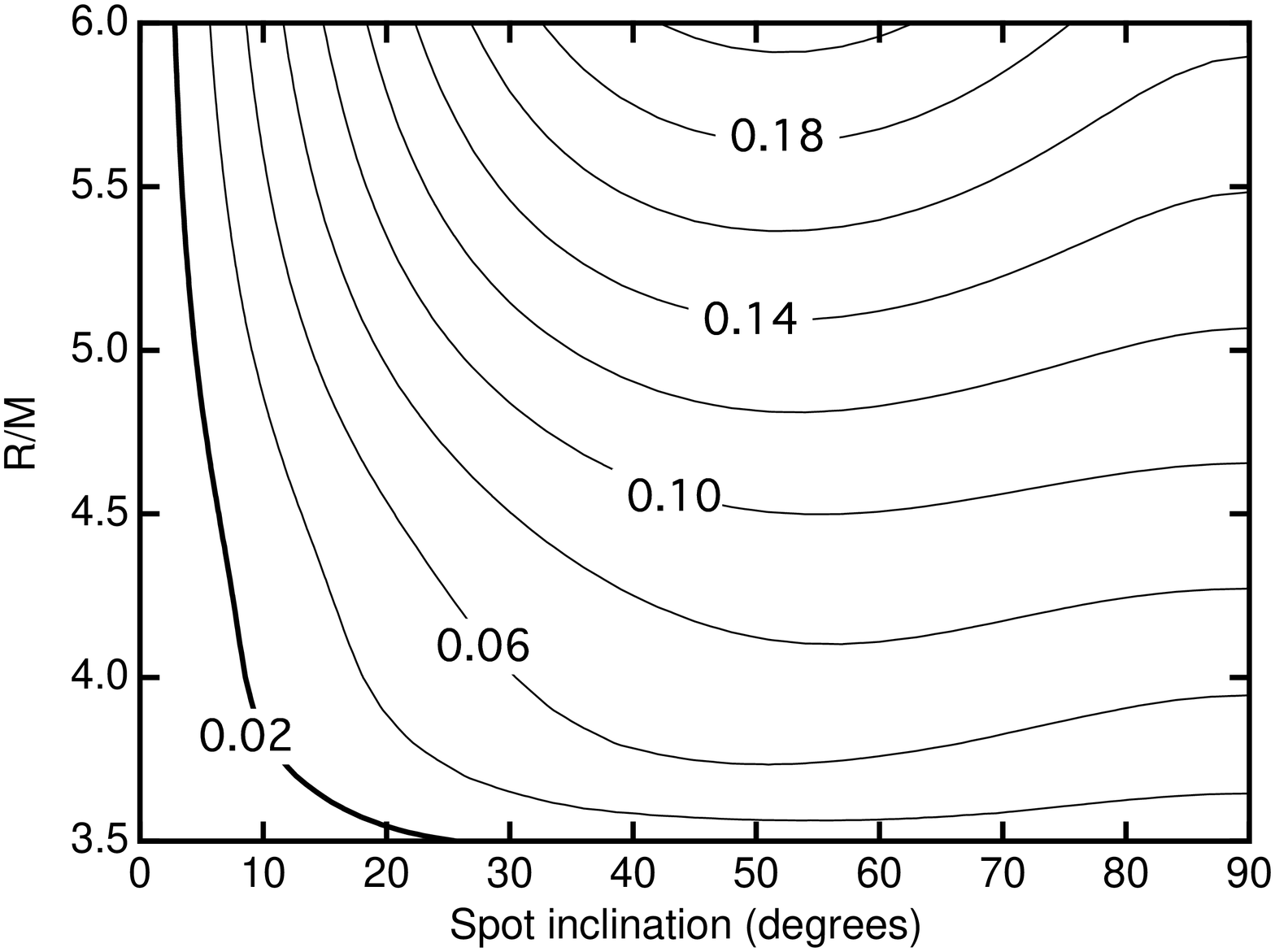}
\caption{
Contour plots of the total fractional rms amplitude produced by
isotropic emission from a single stable spot (left-hand panel) and from
two stable antipodal spots (right-hand panel), as a function of the
stellar radius and the inclination of the primary spot relative to the
spin axis, for a $1.4\,M_\odot$ star with a radius of $5M$ spinning at
$400\,$Hz observed at an inclination of 45$\arcdeg$. The heavier lines
highlight the contours where the fractional flux modulation is 2\%. This
figure demonstrates that compactness alone cannot explain the
low-amplitude oscillations of the \mbox{AMXPs}.
}
\label{fig:amplitudes-compactness}
\end{figure*}

The fractional modulation produced by a given emission pattern is
generally smaller for more a compact star (see, e.g. \citealt{pech83,
stro92}). The reason is that gravitational focusing allows the observer
to see a larger fraction of the surface of such a star, averaging the
contributions of its bright and dark areas and reducing the flux
variation seen as the star turns. Strong gravitational focusing can
increase the fractional modulation seen by some observers if the star is
very compact ($R\la3.5M$).

Even if all \mbox{AMXPs} are relatively compact ($R\sim 4M$), which is
not expected, they would produce fractional modulations larger than the
$\sim\,$1\%--2\% fractional modulations often observed (see Section~\ref
{sec:intro}) unless their emitting regions are close to their stellar
spin axes. Nor can the failure so far to detect accretion-powered
oscillations in some nuclear-powered \mbox{AMXPs} be explained unless
the accretion-powered emission comes from areas very close to the
stellar spin axis. These points are illustrated by
Figure~\ref{fig:amplitudes-compactness}, which shows the fractional
amplitudes produced by isotropic emission from a single spot (left-hand
panel) and from two antipodal spots (right-hand panel) as a function of
$R/M$ and the inclination of the primary spot relative to the spin axis,
for a $1.4M_\odot$ star spinning at 400~Hz observed at an inclination of
45$\arcdeg$.

The left-hand panel of Figure~\ref{fig:amplitudes-compactness} shows
that even if all the \mbox{AMXPs} had radii as small as $4M$, their
fractional modulations could be as small as $\sim\,$2\% for emission
from a single spot only if the spot is within $3\arcdeg$ of the spin
axis. If instead the spot is at a moderate-to-high inclination, their
fractional modulations would be much larger. For example, if the spot is
$45\arcdeg$ from the spin axis, the fractional modulation would be
$\sim\,$25\%.

The right-hand panel of Figure~\ref{fig:amplitudes-compactness} shows
that although the fractional modulation produced by two antipodal spots
is generally smaller than the modulation produced by a single spot,
antipodal spots would have to be within $10\arcdeg$ of the spin axis to
produce a fractional modulation as small as $\sim\,$2\%, even for a star
as compact as $4M$. If instead the spots are $45\arcdeg$, from the spin
axis, the fractional modulation would be $\sim\,$7\%, much larger than
the lowest modulations observed in most \mbox{AMXPs}.

As another example (not plotted here), consider a relatively massive
$2.2\,M_\odot$ star with one or two spots that are not close to the spin
axis, observed at an inclination of 45$\arcdeg$. If there is a single,
isotropically emitting spot of radius of 25$\arcdeg$ at an inclination
of $45\arcdeg$, the fractional modulation would be $\sim\,$30\% if the
star has a radius of $4M$ and 22\% if the star has a radius of $3.2M$.
If instead there are two antipodal, isotropically emitting spots at
inclinations of 45$\arcdeg$ and 135$\arcdeg$, the fractional modulation
would be $\sim\,$11\% if the star has a radius of $4M$ and 9\% if the
star has a radius of $3.2M$. These fractional modulations are much
larger than the lowest modulations observed in most \mbox{AMXPs}. The
fractional modulation is typically even larger for stars that are still
more compact, because of the strong gravitational focusing effect
discussed above.

These results show that even if the \mbox{AMXPs} are compact neutron
stars, their fractional modulation will be much larger than the low
fractional modulations they often produce, unless their emitting areas
are within a few degrees of the spin axis. Hence the low fractional
modulations often observed in the \mbox{AMXPs} cannot be attributed to
high stellar compactness.

\begin{figure*}[t]
\includegraphics[height=.26\textheight]{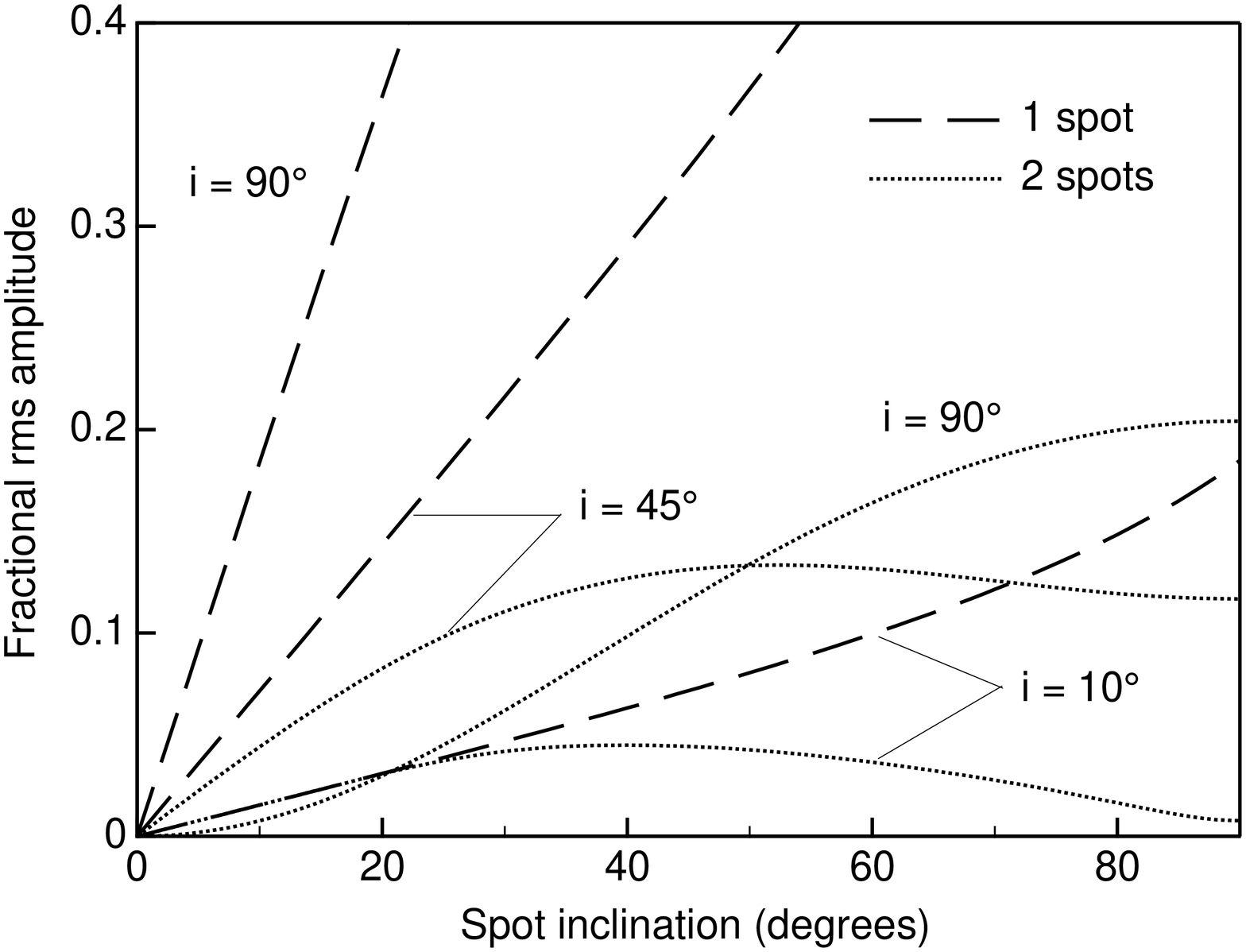}
\includegraphics[height=.26\textheight]{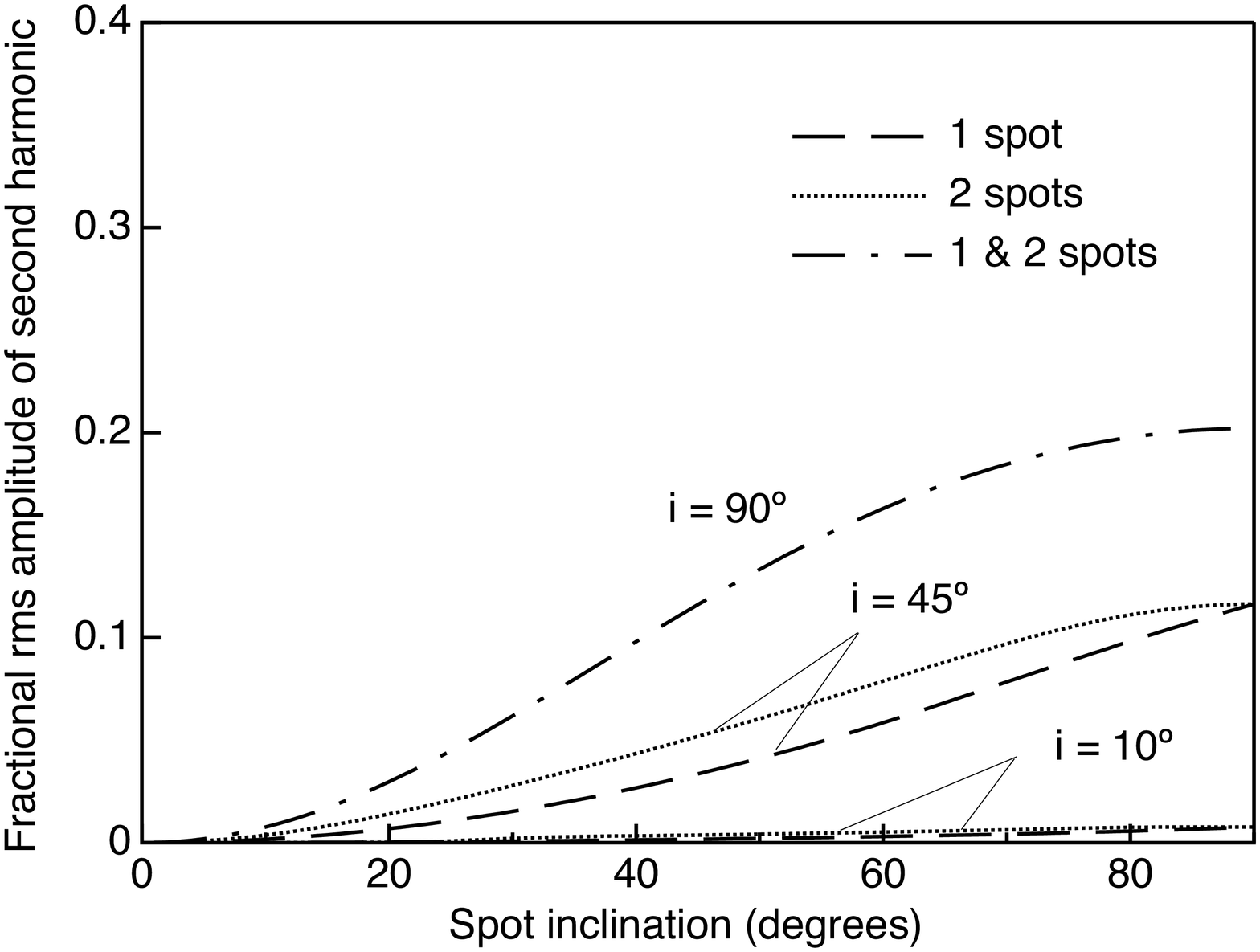}
\caption{
Fractional rms amplitudes of the full bolometric pulse profile (left)
and the second harmonic (first overtone) component of the pulse profile
(right) produced by isotropic emission from a single stable spot and
from two stable antipodal spots, as functions of the inclination of the
primary spot relative to the spin axis, for spot radii of 25$\arcdeg$, a
$1.4\,M_\odot$ star with a radius of $5M$ spinning at $400\,$Hz, and the
indicated observer inclinations $i$. This figure shows that if the
inclination of the emitting region is small, modest changes in its
inclination can produce large changes in the observed amplitude of the
oscillation.
}
\label{fig:amplitudes-inclinations}
\end{figure*}

A further difficulty in attributing the generally low fractional
modulations of the \mbox{AMXPs} to high stellar compactness is that
several of the \mbox{AMXPs} that exhibit fractional modulations
$\sim\,$1\%--2\% at some times exhibit fractional modulations
$\sim\,$15\%--25\% only a few hours or days later. The compactness of
the neutron star cannot change significantly on such short timescales,
and hence some mechanism other than high compactness must be responsible
for the low fractional modulations seen at many times in most
\mbox{AMXPs}.

Our results also show that high stellar compactness cannot be the main
factor responsible for the nondetection of accretion-powered
oscillations in many accreting neutron stars in low-mass X-ray binary
systems, including many stars in which nuclear-powered oscillations have
been detected. The upper limits on the fractional amplitude of any
accretion-powered oscillation produced by these stars are $\la\,$0.5\%
(see Section~\ref {sec:intro}), much less than could be explained by the
effects of high stellar compactness.

\subsection {Dependence on Stellar Spin Rate}
\label {sec:spin-rate}

Most of the results discussed in this section are for stars spinning at
400~Hz. Other things being equal, stars spinning more rapidly will
produce oscillations with larger fractional amplitudes, because their
higher surface velocities will produce larger Doppler boosts and greater
aberration, making their radiation patterns more asymmetric
(\citealt{mill98a, braj00}). Conversely, stars spinning more slowly tend
to produce oscillations with smaller fractional amplitudes. However, the
dependence of the fractional amplitude on the stellar spin rate is weak.
Consequently, the basic conclusions reached in this section are valid
for the full range of \mbox{AMXP} spin rates observed.

\section {Pulse amplitude and phase variations}
\label {sec:variations}

As discussed in Section~\ref {sec:intro}, the fractional amplitudes of
the accretion-powered oscillations of the \mbox{AMXPs} vary by factors
ranging from $\sim\,$2 to $\sim\,$10 on timescales of hours to days. As
was also discussed in Section~\ref {sec:intro}, the phases of the
Fourier components of these oscillations vary by \mbox{$\sim\,$0.1}--0.4
cycles on similar timescales. If interpreted as caused by changes in the
stellar spin rate, the observed phase variations would imply frequency
variations more than a factor of 10 larger than expected for the largest
accretion torques and the smallest inertial moments thought possible for
these stars. In several \mbox{AMXPs}, the phase variations are
correlated with the amplitude variations, for some amplitude ranges. A
successful model of the accretion-powered oscillations of the
\mbox{AMXPs} should provide a consistent explanation of these properties
of the oscillations.

As noted in Section~\ref{sec:intro}, changes in the latitude and
longitude of the emitting area are expected on a wide range of
timescales, from the $\sim\,$0.1~ms dynamical time at the stellar
surface to the $\sim\,$10~d timescales of the variations observed in the
mass accretion rates of the \mbox{AMXPs}. Changes in the latitude
(inclination) of the emitting area primarily affect the amplitudes of
the harmonic components of the pulse profile but also affect their
phases. Modest changes in the latitude of the emitting area can produce
large changes in the oscillation amplitude if the emitting region is
close to the stellar spin axis. Changes in the longitude (stellar
azimuth) of the emitting area shift the phases of the harmonic
components of the pulse but do not affect their amplitudes.

In this section we show that if the emitting area is close to the spin
axis, the fractional amplitude variations seen in all \mbox{AMXPs} and
the larger fractional amplitudes $\sim\,$15\%--20\% occasionally seen in
several of them can be explained by modest variations of the
inclinations of their emitting areas. We show further that if the
emitting area is close to the spin axis and moves in the azimuthal
direction by even a small distance, the phases of the Fourier components
of the pulse will shift by a large amount. These shifts can easily be as
large as \mbox{$\sim\,$0.1}--0.4 cycles. Indeed, if the emitting area
loops the spin axis, the phases of the Fourier components will shift by
more than one cycle. We show that changes in the latitude and longitude
of the emitting area by modest amounts can produce substantial
correlated changes in the amplitudes and phases of the Fourier
components of the oscillation.

\subsection {Pulse Amplitude Variations}
\label {sec:amplitude-variations}

Figure~\ref{fig:amplitudes-inclinations} shows the total fractional rms
amplitude and the fractional rms amplitude of the second harmonic (first
overtone) of the spin frequency for pulses produced by isotropic
emission from a single stable spot and from two stable antipodal spots,
as functions of the inclination of the primary spot relative to the spin
axis. These curves are for spots with angular radii of $25\arcdeg$ on
our reference star.

The results shown in Figure~\ref{fig:amplitudes-inclinations} support
our conclusion in Section~\ref {sec:amplitudes} that the oscillation
amplitudes seen by most observers can be as small as $\sim\,$1\%--2\%
only if the emitting regions are within a few degrees of the spin axis.
If instead the emitting regions were $\ga\,$45$\arcdeg$ from the spin
axis, most observers would see modulation fractions $\ga\,$15\%.

Whether the observer sees emission from a single area or from two
antipodal areas, the total fractional amplitude of the oscillation will
increase if the inclination of an emitting region initially at a low
inclination increases. The reason is that the star's radiation pattern
becomes steadily more asymmetric about the spin axis as the emitting
region moves farther from the spin axis, partly because the emitting
region becomes more asymmetric and partly because the surface velocity
becomes higher.

If the emitting area is initially close to the spin axis, a modest
increase or decrease in its latitude can change the amplitude of the
observed flux oscillation by a substantial factor. For emission from a
single spot, the amplitude increase is steepest for observers at high
inclinations, whereas for emission from two antipodal spots, the
amplitude increase is steepest for observers at moderate inclinations.

These effects are illustrated by the left-hand panel of
Figure~\ref{fig:amplitudes-inclinations}. It shows that for a single
spot initially at $i_{s} = 10\arcdeg$, an increase of the spot
inclination by as little as $\sim\,$5$\arcdeg$ can increase the
fractional amplitude of the oscillations seen by an observer at
45$\arcdeg$ by a factor of 2. For observers at higher inclinations,
the oscillation amplitude increases even more.

A single spot at a high inclination produces oscillations with a high
fractional amplitude for observers at inclinations $\ga\,$60$\arcdeg$
and the amplitude increase only slightly as the inclination increases
further. As an example, for the parameter values assumed in
Figure~\ref{fig:amplitudes-inclinations}, the total fractional amplitude
seen by an observer at $i=45\arcdeg$ increases steadily with increasing
spot inclination, reaching $\sim\,$70\% when $i_s = 90\arcdeg$. In
contrast, the fractional amplitude seen by an observer at
$i\approx90\arcdeg$ (i.e., close to the spin equator) is $\sim\,$85\%
for $i_s=70\arcdeg$ and $\sim\,$90\% for $i_s=90\arcdeg$.

If instead the observer sees radiation from two antipodal spots, the
total fractional amplitude increases with spot inclination for spot
inclinations $\la\,$50$\arcdeg$, though not as steeply as for a single
spot. The fractional amplitude eventually decreases with increasing spot
inclination, unless the observer is located in the spin equator. A
change in the inclination of two antipodal spots by 10$\arcdeg$ can
change the total fractional amplitude seen by an observer at 45$\arcdeg$
by a factor of 2.

The right-hand panel of Figure~\ref{fig:amplitudes-inclinations} shows
that the fractional amplitude of the second harmonic is almost the same
for radiation from one spot or from two antipodal spots, independent of
the inclinations of the observer and the spot(s). Comparison of the
right-hand panel of Figure~\ref{fig:amplitudes-inclinations} with the
left-hand panel shows that the second harmonic dominates the pulse
profile for spot inclinations $\la\,$80$\arcdeg$.

\begin{figure*}[t]
\hspace{30pt}
\includegraphics[height=.27\textheight]{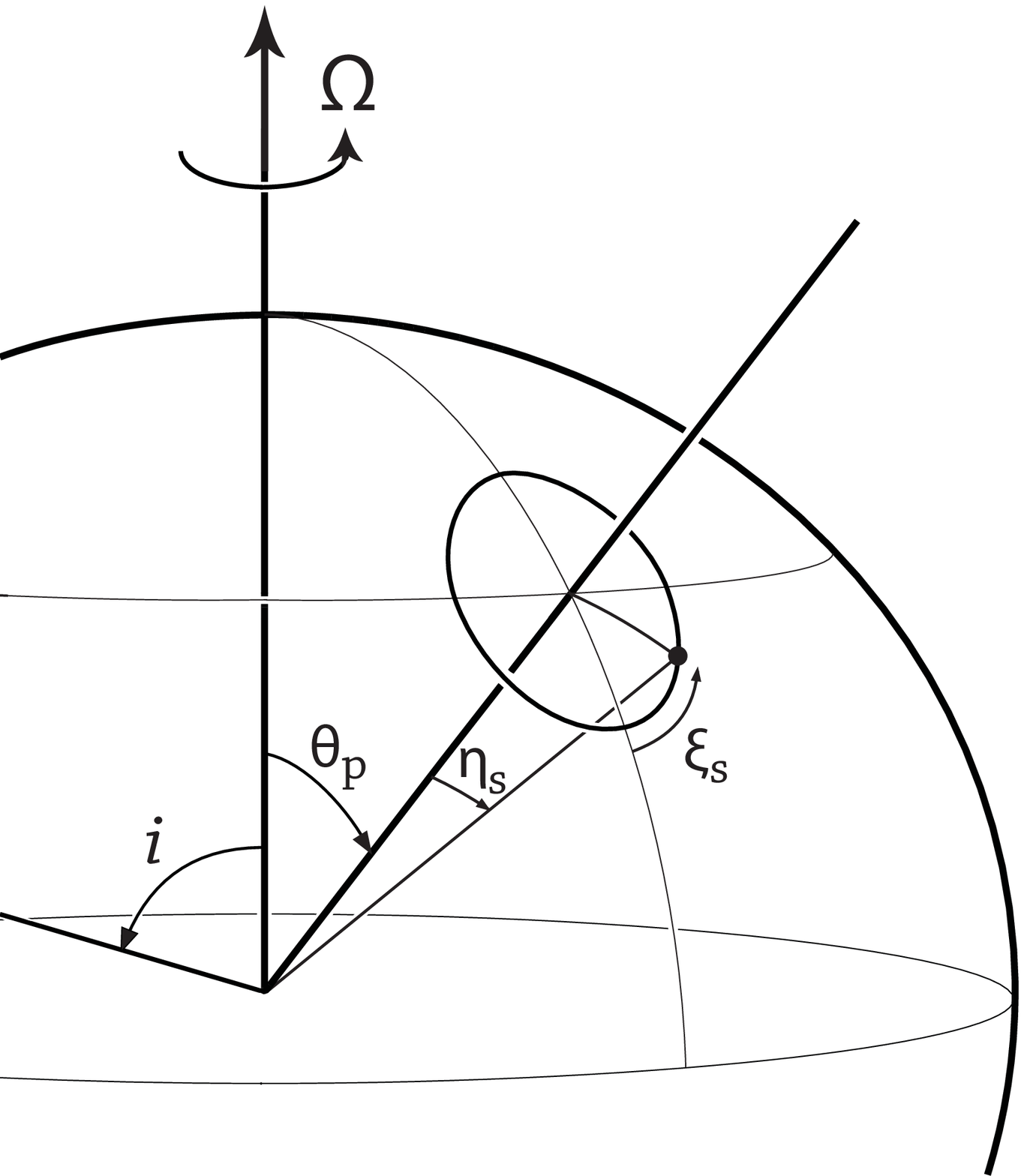}
\hspace{0pt}
\includegraphics[height=.27\textheight]{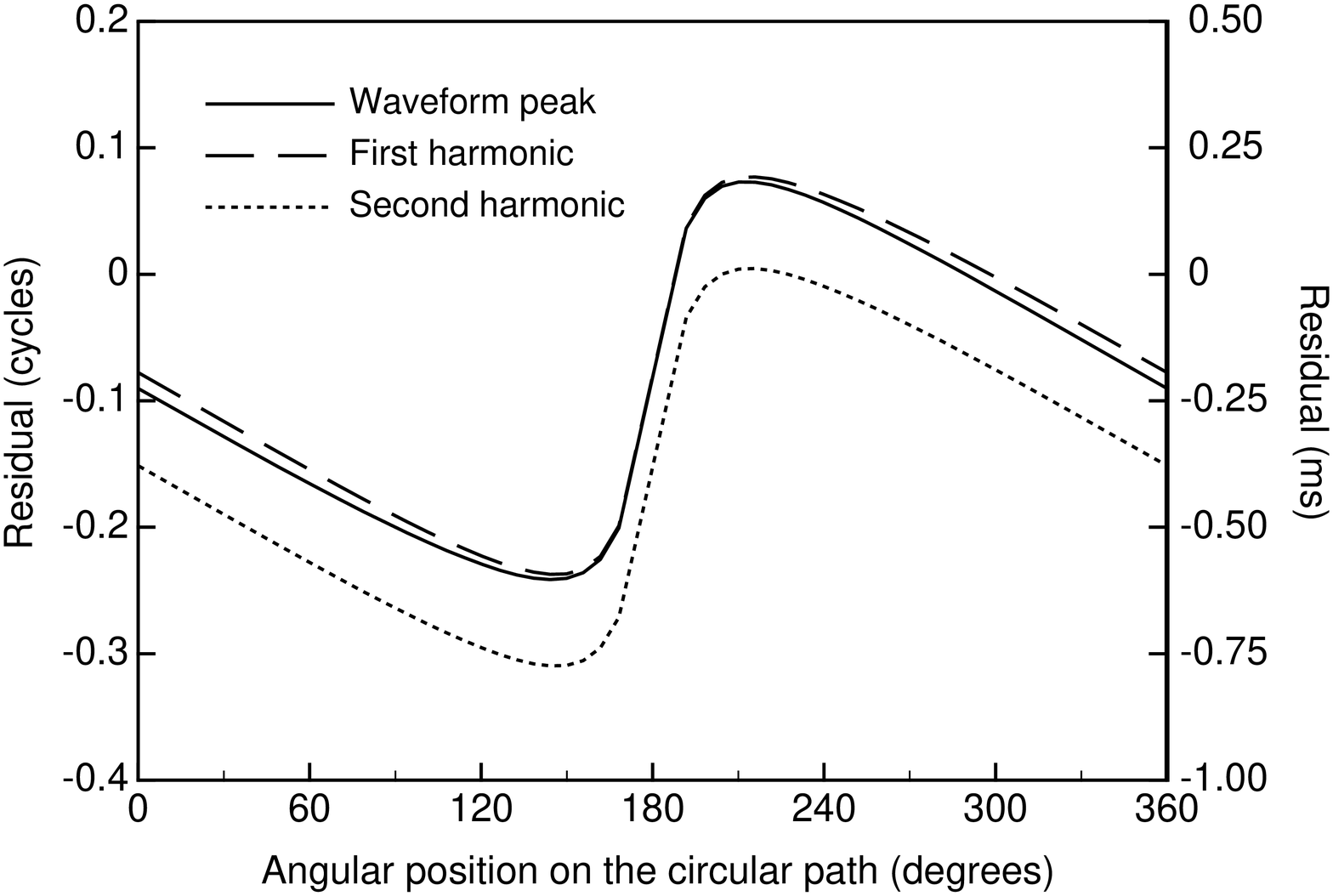}
\vspace{2pt}
\caption{
Left: angles relevant to the path of the emitting spot considered here.
The observer's inclination $i$ is measured from the stellar spin axis
$\Omega$. The path on the stellar surface along which the emitting spot
moves is described in a coordinate system anchored in the star. The
center of the circular path considered here is at colatitude $\theta_p$
and azimuth $\phi_p \equiv 0$ in this coordinate system. The angular
radius of the path is $\eta_s$. The azimuth $\xi_s$ of the spot around
the center of the path is measured from its lowest latitude. Right:
phase and time residuals of the peak, first harmonic (fundamental), and
second harmonic (first overtone) of the observed waveform as functions
of the azimuth $\xi_s$ of a circular spot with a radius of $25\arcdeg$
moving along a circular path like that shown in the left-hand panel. The
phase and time residuals are defined in
Section~\ref{sec:constructing-profiles}. In this example, the emission
from the spot is assumed to be isotropic, $\theta_p = 12\arcdeg$,
$\eta_s = 10\arcdeg$, $i=45\arcdeg$, the star has a mass of
$1.4~M_\odot$ and a radius of $5M$, and the star's spin rate is 400~Hz
measured at infinity.
}
\label{fig:phase-residuals}
\end{figure*}

The accretion-powered oscillations observed in XTE~J1814$-$338 and
XTE~J1807$-$294 occasionally have fractional amplitudes as large as 11\%
and 19\%, respectively, although they are usually much smaller
(\citealt{chun08, patr08, zhan06, chou08}).
Figure~\ref{fig:amplitudes-inclinations} shows that amplitudes this
large are possible for isotropic emission from a single spot even for
relatively small spot inclinations, provided the observer's line of
sight is not too close to the star's spin axis. For example, amplitudes
this large are possible for spot inclinations
$\sim\,$5$\arcdeg$--$15\arcdeg$ if the observer's inclination is
$\ga\,$50$\arcdeg$. A decrease in the spot inclination to
$\sim\,$1$\arcdeg$--$3\arcdeg$ would then reduce the fractional
amplitude to $\sim\,$1\%--3\%, the lowest values observed in these
\mbox{AMXPs}.

If the emitting areas of the \mbox{AMXPs} are typically located at high
stellar latitudes, as in the model discussed here, their pulse
amplitudes are expected to depend on the accretion rate and structure of
the inner disk. These may vary on timescales as short or shorter than
the $\sim\,$0.1~ms dynamical time at the stellar surface up to
timescales as long as the $\sim\,$10~d observed variations of the mass
accretion rate. Consequently, the harmonic amplitudes of \mbox{AMXP}
pulses are expected to vary on these timescales.

This model can be tested on timescales longer than the
$\sim\,$$10^{2}$--$10^{3}$~s integration times required to construct
pulse profiles by searching for correlations between the amplitudes of
the harmonic components of the pulse profile or the variance of the
pulse amplitude and the pulsar's X-ray flux or spectrum. Indeed, the rms
fractional amplitudes of the pulses of XTE~J1814$-$338 appear to be
positively correlated with its mean X-ray flux (\citealt{chun08}).
Methods for testing the model on timescales shorter than the time
required to construct a pulse profile are discussed in Section~\ref
{sec:spot-movements}.

\subsection {Pulse Phase Variations}
\label{sec:phase-variations}

A change in the longitude (stellar azimuth) of the emitting region
alters the arrival time of the pulse. This shifts the measured phases of
all the Fourier components by the same amount, relative to their phases
if the pulse were produced by an emitting area fixed on the surface of a
star rotating at a constant rate. Obviously, the phase of the pulse peak
is also shifted by this amount. If the emitting area is close to the
spin axis, a displacement in the azimuthal direction by even a small
distance can produce large phase shifts.

A change in the latitude (inclination) of the emitting area not only
alters the amplitudes of the Fourier components of the pulse but also
their arrival times (phases). These alterations change the shape and
arrival time of the pulse in a complex way. They occur because a change
in the latitude of the emitting area changes the velocity of the
emitting surface, altering the aberration, Doppler shift, and
propagation time of each photon. The resulting changes in the pulse
shape and arrival time depend on the beaming pattern of the radiation,
the properties of the star, and the inclination of the observer. The
phase shifts caused by changes in the latitude of the emitting area are
expected to be modest, because the surface velocities of neutron star
models with spin frequencies $\la\,$600~Hz are typically a small
fraction of the speed of light.

Figure~\ref{fig:phase-residuals} shows how the phases of the peak of the
pulse and its first and second harmonic components vary as the emitting
spot moves along a circular path on the stellar surface. Motion along
this path changes both the latitude and the longitude of the spot.
Motion in either direction along all or part of such a path may occur if
the main impact area of the accreting matter moves around the magnetic
pole as a result of changes in the structure of the inner disk, changes
that could be caused by variations of the accretion rate onto the star
(see Section~\ref{sec:modeling-emission}).

The left-hand panel of Figure~\ref{fig:phase-residuals} introduces the
angles used to describe the motion of the spot around the path. The
right-hand panel shows the changes in the arrival times and phases of
the peak of the pulse and its Fourier components that occur as the spot
moves around the path shown in the left-hand panel. In this example, the
phases of the peak and the first and second harmonic components vary by
$\sim\,$0.35~cycles, which is comparable to the phase variations seen in
XTE~J1807$-$294 \citep{mark04} and SAX~J1808.4$-$3658
\citep{morg03,hart08} on timescales of hours to days. If the path of the
emitting region approaches or loops the spin axis, even a very small
displacement can shift the phases by more than one cycle. The phase
shifts shown in Figure~\ref{fig:phase-residuals} are typical for
isotropic emission. The phase shifts for Comptonized-emission-beaming
patterns like those proposed by \citet{pout03} are similar, but differ
in detail.

The phase and time residuals shown in Figure~\ref{fig:phase-residuals}
are not zero when the spot is in the plane defined by the spin axis and
the observer ($\xi_s = 0$) for two reasons. First,  when $\xi_s = 0$ the
spot is not directly beneath the observer in this example, so a photon
emitted at this moment will reach the observer later than would a photon
emitted from the point directly beneath the observer, which is the
emission point that defines the zero of the arrival time at the observer
(see Section~\ref{sec:constructing-profiles}). Second, the spinning
motion of the stellar surface produces aberrations, time delays, and
Doppler energy shifts that distort the pulse, causing its harmonic
components to reach the observer at times different from their arrival
times from a nonrotating star. In this particular example these effects
cause the peak of the pulse and its harmonic components to arrive
earlier than they would if the star were not rotating. The phase advance
produced by these effects more than offsets the backward phase shift
produced by the extra travel time.

As the emitting spot moves around the path shown in
Figure~\ref{fig:phase-residuals}, its inclination increases from
2$\arcdeg$ to 22$\arcdeg$. This causes the amplitude of the oscillation
to increase from \mbox{$\sim\,$1\%} when the spot is closest to the
rotation pole to \mbox{$\sim\,$15\%} when it is farthest from the pole.
This variation is comparable to the amplitude variations observed in
XTE~J1814$-$338 and XTE~J1807$-$294 (\citealt{chun08, patr08, zhan06,
chou08}).

As described above, the change in the inclination of the spot as it
moves around the path shown in Figure~\ref{fig:phase-residuals} also
contributes to the phase shifts. In general, an increase in the
inclination of an emitting spot by 20$\arcdeg$ can by itself shift the
phases of the Fourier components of the pulse by amounts ranging from
$\sim\,$0.05 to $\sim\,$0.20~cycles, depending on the beaming pattern,
stellar properties, and viewing direction, and can shift the phases of
different harmonic components by different amounts. These effects are
included in the phase shifts shown in Figure~\ref{fig:phase-residuals}.

These results show that if the emitting areas of the \mbox{AMXPs} are
usually near their spin axes, as in the model discussed here, movement
of the emitting area by modest distances can cause the arrival times of
pulse Fourier components to vary by the large amounts observed, even if
the \mbox{AMXPs} spin at a nearly constant rate.

Emitting areas occur where accreting plasma impacts the stellar surface.
As discussed in Section~\ref {sec:modeling-emission}, the position of
these areas depends on how accreting plasma enters the magnetosphere and
hence on the accretion rate and the structure of the inner disk.
Consequently, pulse phase variations produced by changes in the location
of the emitting areas should be correlated with changes in the
luminosity and spectrum of the pulsar. The properties of the accretion
flow are expected to vary on timescales ranging from as short as
$\sim\,$0.1~ms to as long as $\sim\,$10~d, and the arrival times of
pulse Fourier components should therefore vary on these timescales also.

This model of pulse phase variations can be tested on timescales longer
than the $\sim\,$$10^{2}$--$10^{3}$~s times required to construct a
pulse profile by searching for and studying correlations between the
measured phases of the harmonic components of pulses or the variance of
these phases and the pulsar's X-ray flux or spectrum. The phase
residuals of the first and second harmonic components of the pulses of
XTE~J1814$-$338 appear to be anticorrelated with its X-ray flux
(\citealt{chun08, papi07}), making it a good candidate for this kind of
study. \citet{rigg08} have reported a strong anticorrelation between the
X-ray flux of XTE~J1807$-$294 and the phase residual of the second
harmonic component of its pulse profile. Methods for testing this model
of phase variations on timescales shorter than the time required to
construct a pulse profile are discussed in Section~\ref
{sec:spot-movements}.

\subsection {Correlated Pulse Amplitude and Phase Variations}
\label {sec:correlated-variations}

\begin{figure}[t]
\includegraphics[height=.265\textheight]{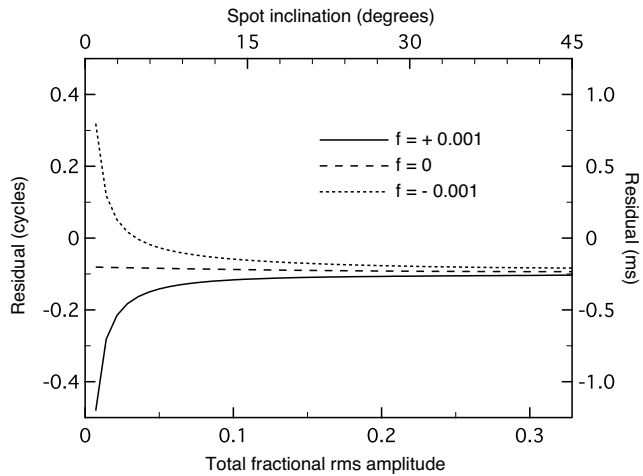}
\caption{
Distribution of pulse arrival times as a function of the pulse amplitude
expected for pulses produced by an emitting area that is close to the
star's spin axis and wanders. In this example the emitting area is
assumed to wander in stellar longitude by the same distance $d\ell$ at
all latitudes. The dotted and solid curves show the expected scatter of
pulse arrival times (right vertical axis) and arrival phases (left
vertical axis) as a function of the pulse amplitude, assuming $f \equiv
d\ell/2\pi R = \pm0.001$. The dashed curve shows the relation between
the arrival time and amplitude of pulses produced by a spot that remains
at the same longitude as its latitude changes. Here the arrival time of
a pulse is taken to be the arrival time of its peak. The phase and time
residuals are defined in Section~\ref{sec:constructing-profiles}. These
results are for a spot with a radius of $25\arcdeg$ on a $1.4\,M_\odot$
star with a radius of $5M$ spinning at $400\,$Hz, observed at an
inclination of $45\arcdeg$. The amplitude range plotted on the
horizontal axis corresponds to spot inclinations from $2\arcdeg$ to
$45\arcdeg$ (compare Figure~\ref{fig:amplitudes-inclinations}). This
figure illustrates the large scatter expected when the fractional
amplitude is small.
}
\label{fig:phase-scatter}
\end{figure}

If the emitting areas of the \mbox{AMXPs} are near their spin axes and
move around with time, the amplitudes and phases of their pulses should
show two types of correlated behavior.

A strong expectation in this model is that the scatter of the pulse
arrival times (i.e., the scatter of the time or phase residuals) should
decrease steeply with increasing pulse amplitude. The reason for this is
that, as shown previously, a given variation in the azimuthal position
of the emitting area produces a much larger phase shift when the
emitting area is very close to the spin axis than when it is farther
away; in contrast, the oscillation amplitude is much smaller when the
emitting area is very close to the spin axis than when it is farther
away.

This effect is illustrated in Figure~\ref{fig:phase-scatter}, which
shows the expected distribution of pulses in the arrival time versus
amplitude plane if they are produced by emission from an area close to
the star's spin axis that wanders over time. In the example shown, the
phase residuals vary by $\sim\,$0.3~cycles when the fractional amplitude
is $\sim\,$0.02 but by only $\sim\,$0.03~cycles when the fractional
amplitude is $\sim\,$0.1. The pulse phase residuals seen in several
\mbox{AMXPs} (see, e.g., XTE~J1807$-$294: \citealt{patr08}; SAX
J1748.9-2021: \citealt{patr08}; and IGR~J00291$+$5934: \citealt{patr08})
have distributions similar to that shown in
Figure~\ref{fig:phase-scatter}. These distributions are consistent with
the model discussed here if the emitting areas of these \mbox{AMXPs}
wander in the azimuthal direction by distances equal to
$\sim\,$$10^{-2}$--$10^{-3}$ of the stellar circumference.

A second expectation in the moving spot model discussed here is that the
pulse arrival time or phase residuals are likely to form a track in the
phase-residual versus pulse-amplitude plane, especially if the change in
the pulse amplitude is large. Such a track will be formed if the
emitting areas where the accreting matter impacts the stellar surface
tend to move along a favored path in stellar latitude and longitude as
the accretion rate and the structure of the inner disk change. This is
expected because models of the flow of accreting matter from the inner
disk to the stellar surface predict that matter will be guided along
particular but different magnetic flux tubes as the accretion rate and
the structure of the inner disk vary (see Section~\ref
{sec:modeling-emission}).

Our computations show that motion of the emitting area along a path that
changes its latitude but not its longitude can shift the phases of the
first and second harmonics by at least 0.15~cycles and by different
amounts. Motion of the emitting area along a path that instead changes
its longitude but not its latitude can shift the phases of the first and
second harmonics and the pulse peak by much larger amounts, if the
emitting area is near the spin axis, but in this case the phase shifts
will be the same for all Fourier components. Only if the shift in pulse
phase produced by the change in the longitude of the emitting area
exactly compensates for the shift produced by the change in the latitude
of the emitting area will there be no trend in the pulse phase residuals
with increasing pulse amplitude.

No track may be discernible in the phase-residual versus pulse-amplitude
plane when the pulse amplitude is small because, as has just been
discussed, the pulse phase is very sensitive to the position of the
emitting area when the area is near the spin axis, which is where it is
expected to be when the pulse amplitude is small. If, however, the
emitting area moves well away from the spin axis, which in the moving
spot model produces a substantial increase in the pulse amplitude, a
track may become apparent. Tracks are observed in the data on several
\mbox{AMXPs} when the pulse amplitude is $\ga\,$3\% (\citealt{patr08}).
The phase of the first harmonic component of the pulse is correlated
with the amplitude of the pulse in IGR~J00291$+$5934 and XTE~J1751$-$305
but anticorrelated in XTE~J1807$-$294 and XTE~J1814$-$338.

The path on the stellar surface along which the emitting area moves as
the accretion rate changes depends on the details of the accretion flow
from the inner disk to the stellar surface that cannot yet be determined
from first principles. However, the path along which the emitting area
moves in the moving spot model can be inferred from simultaneous
measurements of the phases and amplitudes of pulses.

Similar shifts in the phases of all Fourier components are consistent
with movement of an emitting area of roughly fixed size, shape, and
radiation-beaming pattern along a path that causes a substantial change
in its longitude as its latitude decreases. This behavior is observed in
XTE~J1814$-$338 (\citealt{papi07, chun08, patr08}). If instead the phase
residuals of different Fourier components of the pulse evolve very
differently with increasing pulse amplitude, this is an indication that
the emitting area is moving along a path that produces very little
change in the area's longitude and/or a substantial change in its shape
or the radiation-beaming pattern. This behavior is observed in
XTE~J1807$-$294 (\citealt{patr08}).

\subsection{Accretion- and Nuclear-Powered Oscillations}
\label{sec:accretion-nuclear}

The close agreement of the pulse profiles and phases of the accretion-
and nuclear-powered (X-ray burst) oscillations observed in
SAX~J1808.4$-$3658 \citep{chak03} and XTE~J1814$-$338 \citep{stro03}
strongly suggests that in these stars, both types of oscillation are
produced by X-ray emission from nearly the same area on the stellar
surface. If this is so, it implies that in these \mbox{AMXPs}
thermonuclear burning is concentrated near the magnetic poles onto which
accreting matter is falling. It also implies that long-term variations
in the phase residuals of the two types of oscillations should track one
another in these pulsars and should also be correlated with variations
in the X-ray flux and spectrum, because both types of variations are
produced by changes in the accretion flow through the inner disk.

This interpretation can be tested by comparing the observed variations
of the amplitudes and phases of the Fourier components of the accretion-
and nuclear-powered oscillations with the variations predicted by this
model. If this interpretation proves correct, the locations and
movements of the emitting areas can be determined from the observed
phase and amplitude variations.

In Section~\ref {sec:pole-movement}, we emphasize that a mechanism that
drives a star's magnetic poles toward its spin axis can greatly reduce
the dipole component of the star's magnetic field without reducing
significantly its strength. The final configuration of the magnetic
field produced by this mechanism could have a total strength
$\sim\,$$10^{11}$--$10^{12}$G, but a dipole component
$\sim\,$$10^{8}$~G, consistent with the dipole moments inferred from the
X-ray emission and spin evolution of \mbox{AMXPs} and millisecond radio
pulsars (\citealt{lamb08a}).

Surface magnetic fields $\sim\,$10$^{11}$--10$^{12}$~G are strong enough
to laterally confine accreting matter within the surface layers of the
neutron star, creating an accumulation point for fuel for thermonuclear
bursts and causing thermonuclear burning to be concentrated near the
star's magnetic poles (\citealt{woos82}; see also \citealt{mela01} and
references therein). They are also strong enough to produce
strong-magnetic-field features in the keV X-ray spectra of \mbox{AMXPs}
(see \citealt {mesz92}). If this picture is correct, such features are
more likely to be detected in AMXPs such as SAX~J1808.4$-$3658 and
XTE~J1814$-$338 that produce nuclear-powered oscillations that are
phase-synchronized (or nearly synchronized) with their accretion-powered
oscillations than in other AMXPs.

\subsection{Effects of Rapid Spot Movements}
\label{sec:spot-movements}

Our results show that motion of the emitting area on the stellar surface
generally changes both the amplitudes and the phases of the Fourier
components of the pulse profile. As noted in Section~\ref
{sec:modeling-emission}, the position of the emitting area is expected
to reflect the accretion rate and structure of the inner disk, and is
therefore expected to change on timescales at least as short as the
$\sim\,$0.1~ms dynamical time near the neutron star. Pulse phase and
shape variations are therefore expected on this timescale.

The observable effects of changes in the position of the emitting area
on timescales longer than the $\sim\,$$10^{2}$--$10^{3}$~s integration
times required to construct a pulse profile have been discussed in
previous sections. Here we discuss the expected effects of fluctuations
in the position of the emitting area on timescales shorter than the time
required to construct a pulse profile.

Changes in the position of the emitting area on timescales shorter than
the time required to construct a pulse profile will appear as unresolved
amplitude and phase noise (see~\citealt{lamb85}). This noise will reduce
the apparent fractional amplitudes of the oscillations at the spin
frequency and its harmonics, making them appear weaker than in the plots
presented in this paper, which assume stable emitting spots fixed on the
stellar surface. These pulse phase and amplitude fluctuations are likely
to be greater when the emitting area is near the spin axis, because
there a displacement by a given distance produces a larger change in the
pulse phase and, for many geometries, in the pulse amplitude. This trend
will tend to reduce further the apparent pulse amplitude when the
emitting area is close to the spin axis, making it appear even smaller
relative to the pulse amplitude when the emitting area is further from
the spin axis than is shown in the plots presented earlier in this
section.

Pulse shape fluctuations caused by rapid changes in the position of the
emitting region will appear as background noise in excess of the Poisson
counting noise. This noise can in principle be detected, especially
because its strength is expected to be anticorrelated with the pulse
amplitude and to depend in a systematic way on the X-ray flux and
spectrum of the pulsar. Detection of more excess noise when the apparent
pulse amplitude is smaller would support the moving spot model of pulse
phase and amplitude variations.

\subsection {Undetected and Intermittent Pulsations}
\label {sec:weak-pulsations}

The results presented in Section~\ref{sec:amplitudes} show that if the
emitting area is very close to the spin axis and remains there, the
amplitude of X-ray oscillations at the stellar spin frequency or its
overtones may be so low that they are undetectable. In addition, rapid
variations in the shape and phase of pulses are expected to be stronger
when the emitting area is very close to the spin axis. The noise
produced by these fluctuations may---in combination with other effects,
such as reduction of the modulation fraction by scattering in
circumstellar gas---further reduce the detectability of
accretion-powered oscillations in neutron stars with millisecond spin
periods~(\citealt{lamb85, mill00}). Taken together, these effects may
help explain the nondetection of accretion-powered oscillations in some
accreting neutron stars in which nuclear-powered oscillations have been
detected.

Location of the X-ray emitting areas close to the spin axis also
suggests a natural explanation for the sudden appearance and
disappearance of accretion-powered oscillations observed in the
``intermittent'' \mbox{AMXPs} (see \citealt{lamb08b, bout08b}). A change
in the accretion flow within the magnetosphere can suddenly channel gas
to the stellar surface farther from the spin axis, causing the centroid
of the emitting area to move away from the axis. This will increase the
amplitude of the oscillation, potentially making a previously
undetectable oscillation detectable. Conversely, a change in the
accretion flow that suddenly channels accreting gas closer to the spin
axis could make a detectable oscillation undetectable. As noted
by~\citet{lamb08b}, this idea can be tested by studying short-term
variations in the amplitudes and phases of the harmonic components of
the pulses.

\section {Discussion}
\label {sec:discussion}

The results presented in previous sections show that a model of
\mbox{AMXPs} in which the X-ray--emitting areas are close to the spin
axis and move around on the stellar surface can explain many of their
properties. Emitting areas close to the spin axis are to be expected if
the magnetic poles of the \mbox{AMXPs} are close to their spin axes,
causing accreting gas to be channeled there. In this section, we first
discuss mechanisms that may cause the magnetic poles of \mbox{AMXPs} to
be close to their spin axes. We then point out that this picture of
magnetic field evolution may also explain why \mbox{AMXPs} in which
accretion-powered oscillations have been detected are transient X-ray
sources. We also discuss several observational tests of this model.

\subsection {Movement of Magnetic Poles Toward the Spin Axis}
\label {sec:pole-movement}

The neutron vortices in the fluid core of a spinning neutron star are
expected to move radially inward if the star is spun up. This inward
vortex motion is expected to drag the magnetic flux tubes in the fluid
core of the star toward the star's spin axis~(\citealt{srin90,
rude91}).\footnote{It has also been argued~\citep{srin90} that a neutron
star's magnetic field will be greatly reduced earlier in its evolution
if it is spun down by a large factor, causing outward moving neutron
vortices to force its magnetic field into the crust near the rotation
equator, where it can be dissipated.} Some of the relevant physics is
not well understood and many details remain to be explored, but this
process is expected to squeeze the magnetic flux of accreting pulsars
toward their spin axes as they are spun up to short periods (recycled).
As a result, the magnetic poles of recycled millisecond pulsars are
expected to be very close to their spin axes~(\citealt{chen93, chen98}).
The magnetic poles of recycled pulsars could also end up close to their
spin axes if the accretion spin-up torque has a component that tends to
align the star's magnetic field with its spin axis as it spins up.
Motion of the magnetic poles toward the spin axis will be facilitated by
diffusion of magnetic flux through the crust and/or fracture and
fragmentation of the crust~(\citealt{rude91}).

If the star's north and south magnetic poles are in opposite rotation
hemispheres when spin-up begins, the inward motion of vortices will drag
them toward opposite spin poles. If instead both poles are in the same
rotation hemisphere when spin-up begins, they will be dragged toward the
same spin pole. In either case, the strength of the magnetic field at
the stellar surface will be orders of magnitude larger than the strength
of its dipole component, as we now explain.

An initial magnetic field $M_1$ that is approximately uniform over a
stellar hemisphere of radius $R$ has a magnetic moment $\mu_1 \approx
M_1 R^3$. The magnetic flux threading the fluid core is conserved as the
magnetic field is squeezed. Suppose the inward moving vortices squeeze
the flux into a tube of radius $a \approx R/30$. This will produce a
squeezed magnetic field of strength $M_2 \approx (R/a)^2 M_1 \approx
10^3 M_1$.

If the star's north and south magnetic poles are forced toward opposite
spin poles, the dipole moment of the squeezed magnetic field will be
aligned with the star's spin axis and will have a magnitude $\mu_2
\approx M_2 R a^2 \approx \mu_1$, corresponding to a surface magnetic
field with a dipole component $M_2({\rm dipole}) \approx M_1$. The full
strength $M_2$ of the surface magnetic field will therefore be $\approx
(R/a)^2 \approx 10^3$ times larger than the strength $M_2({\rm dipole})$
of its dipole component.

If instead the star's north and south magnetic poles are both forced
toward the same spin pole, the dipole moment of the squeezed magnetic
field will be orthogonal to the spin axis and hence the star will be an
``orthogonal rotator'', even though both magnetic poles are near a spin
pole rather than near the spin equator. In this case the dipole moment
will have a magnitude $\mu_2 \approx M_2 L a^2 \approx (L/R)\mu_1$,
where $L$ is the final separation between the north and south magnetic
poles, corresponding to a surface magnetic field with a dipole component
$M_2({\rm dipole}) \approx (L/R)M_1$. Thus if $L \approx 0.1 R$, the
dipole moment of the squeezed field will be about 10 times smaller than
the dipole moment of the original field and the full strength $M_2$ of
the surface magnetic field will be $\approx (R/L)(R/a)^2 \approx 10^4$
times larger than the strength $M_2({\rm dipole})$ of its dipole
component.

Regardless of whether the star's north and south magnetic poles are
forced toward opposite spin poles, creating an aligned rotator, or
toward the same spin pole, creating an orthogonal rotator, both magnetic
poles will end up very close to the spin axis. Accreting gas that is
channeled along magnetic field lines will therefore impact the stellar
surface close to the spin axis and accretion-powered X-ray emission will
come primarily from regions near the spin axis.

Note that if a pulsar's north and south magnetic poles have both been
forced close to the same spin axis, the flow of accreting matter could
impact the surface in a pattern with predominantly either one-fold or
two-fold symmetry relative to the spin axis. In this accretion geometry,
even a small change in the properties of the flow through the inner disk
could change the dominant symmetry of the emitting area from one-fold to
two-fold or vice versa, thereby changing the relative strengths of the
first and second harmonics in the pulse profile.

Another consequence of this picture of magnetic field evolution is that
both accretion-powered (X-ray) and rotation-powered (radio) millisecond
pulsars may have surface magnetic fields $\sim\,$10$^{11}$--10$^{12}$~G.
This is $\sim\,$10$^3$--10$^4$ times stronger than the
$\sim\,$10$^{8}$--10$^{9}$~G dipole components inferred from the
magnetospheric radii of \mbox{AMXPs} (see, e.g., \citealt{psal99,
lamb08a}), the spin-down of SAX~J1808.4$-$3658 in quiescence
(\citealt{hart08}), and the spin-down rates of the rotation-powered
millisecond radio pulsars (see \citealt{lamb08a}). Magnetic fields this
strong are strong enough to create an accumulation point for fuel for
thermonuclear bursts and produce strong-magnetic-field features in the
keV X-ray spectra of \mbox{AMXPs} (see
Section~\ref{sec:accretion-nuclear}).

\subsection {Why \mbox{AMXPs} are Transients}
\label {sec:transients}

The picture of the X-ray emission and magnetic field evolution of
\mbox{AMXPs} that we have outlined here suggests a possible explanation
for why all \mbox{AMXPs} found so far are transients. These systems have
very low long-term average mass transfer rates, but binary modeling
suggests that the mass transfer rates were higher in the past (see,
e.g., \citealt{bild01} for a discussion in the context of
SAX~J1808.4$-$3658). If stars such as these are initially spun up to
high spin rates, so that their magnetic poles are forced very close to
their spin axes, they would appear similar to the accreting neutron
stars in low-mass X-ray binary systems in which accretion-powered
oscillations have not been detected. When their accretion rates later
decrease, and magnetic dipole and other braking torques cause them to
spin down, their magnetic poles will be forced away from the rotation
axis, and their accretion-powered oscillations will become detectable.

If this explanation of the transient nature of the \mbox{AMXPs} is
correct, they should be spinning down on long timescales, as seems to be
the case (see, e.g., \citealt{hart08}), and the amplitudes and phases
the harmonic components of their waveforms should evolve in a correlated
way as their magnetic poles move away from their spin axes (see
Section~\ref{sec:variations}).

\subsection {Comparison with the Properties of \mbox{MRPs}}
\label {sec:rpmsps}

Rotation-powered millisecond radio pulsars (\mbox{MRPs}) are thought to
be the offspring of \mbox{AMXPs} and therefore should have magnetic
field geometries similar to those of the \mbox{AMXPs}, except that the
low rate of accretion at the end of the accretion phase and magnetic
dipole braking toward the end of the accretion phase and afterward will
tend to spin them back down, forcing their magnetic poles away from
their spin axes.

Although the radio emission properties of most \mbox{MRPs} provide
little clear evidence about the locations of their magnetic poles
\citep[see][]{manc04}, there are indications from their radio emission
that the dipole moments of a substantial number of the most rapidly
spinning \mbox{MRPs} in the galactic disk are nearly aligned or nearly
orthogonal to their spin axes (\citealt{chen98}; see also
\citealt{chen93}). As explained above, either orientation is consistent
with their magnetic poles being driven close to their spin axes by
neutron vortex motion during spin-up as \mbox{AMXPs}.

Analysis and modeling of the waveforms of the thermal X-ray emission
from \mbox{MRPs} may provide better constraints on their magnetic field
geometries. Recent modeling of the high (30\% to 50\% rms) amplitude
X-ray oscillations observed in three nearby \mbox{MRPs} using
Comptonized emission and two antipodal or nearly antipodal hot spots is
consistent with their emitting regions being far from their spin axes
\citep{bogd07, bogd08}. We note that all three pulsars have relatively
low ($\sim\,$150~Hz--200~Hz) spin rates and may therefore have been spun
down by a factor $\sim 3$ from their maximum spin frequencies after
spin-up. If so, their magnetic poles may have been forced away from
their spin axes by a similar factor. This could allow consistency
between the $\sim 40\arcdeg$--$60\arcdeg$ inclinations inferred by
\citeauthor{bogd07} for these three \mbox{MRPs} and the $\la 20^\circ$
inclinations typically required in the model of \mbox{AMXPs} described
here. We also note that only \mbox{MRPs} that produce high-amplitude
X-ray oscillations are currently detectable as oscillators, so the ones
that are detected may have magnetic inclinations larger than is typical.
More investigation is needed.

\subsection {Other Observational Tests}
\label {sec:tests}

In the nearly aligned moving spot model of \mbox{AMXP} X-ray emission,
the properties of X-ray pulses (e.g., their amplitudes, harmonic
content, and arrival times) should be functions of the pulsar's X-ray
luminosity and spectrum. The reason is that the properties of pulses are
determined by the location of the emitting area, which depends on the
rate and structure of the accretion flow through the inner disk. The
X-ray luminosity and spectrum of the pulsar will also depend on the
accretion flow through the inner disk. Hence the properties of pulses
should be correlated with the X-ray luminosity and spectrum.

Changes in the location of the emitting area on the stellar surface on
timescales longer than the $\sim\,$$10^{2}$--$10^{3}$~s intervals
required to construct stable pulse profiles will produce correlated
changes in the amplitudes, harmonic content, and arrival times of these
profiles.

In the nearly aligned moving spot model, the arrival times of pulses
with low amplitudes are expected to fluctuate much more than the arrival
times of pulses with high amplitudes (see
Figure~\ref{fig:phase-scatter}). The reason is that pulse amplitudes are
lower when the emitting area is closer to the spin axis, where small
variations in the position of the emitting area can produce large
variations in the pulse arrival time.

A second expectation is that pulse phase residuals will form a track in
the phase-residual versus pulse-amplitude plane. This will be the case
if the area where accreting matter impacts the stellar surface and
X-rays are emitted tends to move along a favored path on the stellar
surface as the accretion flow through the inner disk changes. The
location of the emitting area on its favored path will depend on the
accretion flow through the inner disk and will in turn determine the
position of pulses in the phase-residual versus pulse-amplitude plane.
Changes in the accretion flow through the inner disk will therefore
cause correlated changes in the phases and amplitudes of pulses,
producing a track when these are plotted in the phase-residual versus
pulse-amplitude plane. Such tracks are likely to be more evident if the
range of the pulse amplitude variation is large.

The pulses of IGR~J00291$+$5934, XTE~J1751$-$305, XTE~J1807$-$294, and
XTE~J1814$-$338 seem to form track-like patterns in the phase-residual
versus pulse-amplitude plane when the pulse amplitude is $\ga\,$5\% (see
Section~\ref{sec:correlated-variations}). An important expectation of
the moving spot model is that the position of pulses along such a track
should be correlated with the X-ray luminosity and spectrum of the
pulsar.

Another check of the moving spot model is possible if the nearly
identical pulse profiles and phases of the accretion- and
nuclear-powered X-ray oscillations of XTE~J1814$-$338 and
SAX~J1808.4$-$3658 are produced by emission from the same area on the
stellar surface, as suggested in Section~\ref{sec:accretion-nuclear}. If
this is the case, the model predicts that longer-term (days--weeks)
variations of the phase residuals of the accretion- and nuclear-powered
oscillations should be correlated with one another and with longer-term
variations of the pulsar's X-ray luminosity and spectrum, because all
three variations are expected to be related to changes in the accretion
flow through the inner disk. Evidence of this behavior would support the
moving spot model.

The location of the emitting area on the stellar surface is likely to
change on timescales as short as the $\sim\,$1~ms dynamical timescale in
the inner disk. These rapid movements of the emitting area will cause
rapid variations of the observed X-ray flux. Flux variations on
timescales shorter than the $\sim\,$$10^{2}$--$10^{3}$~s intervals
required to construct stable pulse profiles are observable as noise in
excess of the photon counting noise. In the moving spot model, the
motion of the emitting area that produces this excess noise also affects
the amplitudes, harmonic content, and arrival times of the X-ray pulses,
and is related to the X-ray luminosity and spectrum. Hence the strength
of the excess noise produced by motion of the emitting area should be
correlated with these other properties of the \mbox{AMXP}.

A strong expectation in the nearly aligned moving spot model is that the
component of the excess noise produced by motion of the emitting area
should be stronger when the pulse amplitude is smaller and weaker when
the pulse amplitude is larger (see
Section~\ref{sec:correlated-variations}).

One way to focus on the strength of the noise produced by movement of
the emitting area would be to filter the X-ray flux data to remove
variations on timescales shorter than, say, 0.05~s. This would remove
the variability associated with oscillations at the pulsar spin
frequency and any high-frequency \mbox{QPOs}. The remaining total
variability would then be correlated with the other properties of the
\mbox{AMXP} listed above. Detection of greater variability when the
observed pulse amplitude is smaller would support the nearly aligned
moving spot model.

If \mbox{AMXPs} do have surface magnetic fields as strong as
$\sim\,$$10^{11}$--$10^{12}$~G, as suggested in
Section~\ref{sec:pole-movement}, their keV spectra may show
strong-magnetic-field features. Such features may include increased flux
below the fundamental cyclotron resonance energy, where the magnetic
field suppresses the opacity of surface layers (see, e.g., \citealt
{pavl75}); resonance scattering features produced by the fundamental
cyclotron resonance and its overtones (see, e.g., \citealt {mesz85});
and changes in the spectrum at the higher energies where the cyclotron
scattering becomes unimportant.

Another implication of the existence of \mbox{AMXPs} with surface
magnetic fields as strong as $\sim\,$$10^{11}$--$10^{12}$~G is that the
rotation-powered millisecond pulsars that are their progeny should have
surface magnetic fields this strong, even though their dipole fields are
$\sim\,$$10^{8}$--$10^{9}$~G. Magnetic fields this strong can be
expected to affect particle acceleration and $\gamma$-ray emission by
these pulsars.

\section{Conclusions}
\label{sec:conclusions}

In previous sections we have explored in some detail the nearly aligned
moving spot model of \mbox{AMXP} X-ray emission. In this model the X-ray
emitting regions are close to the stellar spin axis and move around on
the stellar surface with time. Here we list our principal conclusions.

\textit {Pulse amplitudes and shapes}. In
Section~\ref{sec:spot-inclination}, we investigated the amplitudes and
shapes of the pulses produced by emitting spots on the stellar surface
as a function of their inclination to the spin axis, for several X-ray
beaming patterns and a range of stellar masses, compactnesses, and spin
rates. We found that emitting areas on or near the stellar surface can
produce fractional amplitudes as low as the 1\%--2\% values often
observed only if they are located within a few degrees of the stellar
spin axis. Regions near the spin axis also naturally produce nearly
sinusoidal pulse profiles.

We explored effect of spot size on pulse amplitude in
Section~\ref{sec:spot-size}. We found that although the pulse amplitudes
produced by large emitting areas tend to be smaller, this effect is
weak. Unless almost the entire surface of the star is uniformly
emitting, even large spots produce pulse amplitudes greater than those
observed in the \mbox{AMXPs}, unless the spots are centered close to the
spin axis.

In Section~\ref{sec:compactness}, we studied the effect of stellar
compactness on pulse amplitudes. Although the pulse amplitudes produced
by very compact neutron stars tend to be smaller than the amplitudes
produced by less compact stars, we found that this effect is too weak to
explain by itself pulse amplitudes as small as those observed in the
\mbox{AMXPs}. Stellar compactness clearly cannot explain why the
fractional amplitudes of several \mbox{AMXPs} are $\sim\,$1\%--2\% at
some times but $\sim\,$15\%--25\% a few hours or days later, because the
stellar compactness cannot change on such short timescales.

These results show that emission from the stellar surface can explain
the low amplitudes and nearly sinusoidal waveforms typically observed in
\mbox{AMXPs} only if the emitting areas are located close to the stellar
spin axis.

\textit {Variability of pulse amplitudes, shapes, and arrival times}. In
Section~\ref{sec:amplitude-variations}, we investigated the amplitude
changes that can be produced by motion of the emitting area on the
stellar surface. We found that if the emitting area is close to the spin
axis, even a small change in the latitude of the area can change the
oscillation amplitude by a substantial factor. For example, changes in
the inclination of the emitting area by $\la\,$10$\arcdeg$ can explain
the amplitude variations seen in the \mbox{AMXPs} and the relatively
large fractional amplitudes $\sim\,$15\%--20\% occasionally seen in some
of them.

In Section~\ref{sec:phase-variations}, we studied the changes in the
arrival times (phases) of the harmonic components of the pulse caused
when the emitting area moves around on the stellar surface. We found
that changes in the latitude of the emitting area can shift the phases
of the first and second harmonics by at least 0.15~cycles and by
different amounts. We showed that if the emitting area is close to the
spin axis and moves in the azimuthal direction by even a small distance,
the phases of the first and second harmonics can easily shift by as much
as \mbox{$\sim\,$0.1}--0.4 cycles. If the emitting area loops the spin
axis, the phases of the Fourier components will shift by more than one
cycle.

Motion of the emitting area on the stellar surface generally produces
both amplitude and phase variations. As discussed in
Section~\ref{sec:model}, the position of the emitting area is expected
to reflect the accretion rate and structure of the inner disk, and is
therefore expected to change on timescales at least as short as the
$\sim\,$0.1~ms dynamical time at the stellar surface and as long as the
$\sim\,$10~d timescale of the variations observed in the mass accretion
rate. Changes in the position of the emitting area on timescales longer
than the $\sim\,$$10^{2}$--$10^{3}$~s integration times required to
construct a pulse profile will produce changes in the apparent pulse
amplitude and phase.

These results show that if the emitting area is close to the spin axis,
modest changes in its location can explain the rapidly varying harmonic
amplitudes and phases of the \mbox{AMXPs}.

\textit {Correlated amplitude and phase variations}. We showed in
Section~\ref{sec:correlated-variations} that changes in the latitude and
longitude of the emitting area tend to produce correlated changes in the
amplitudes and phases of the harmonic components of the pulse. A strong
expectation in the nearly aligned moving spot model is that the scatter
in the pulse arrival times (i.e., the pulse time or phase residuals)
should decrease steeply with increasing pulse amplitude. The residuals
of several \mbox{AMXPs}, including XTE~J1807$-$294, SAX~J1748.9-2021,
and IGR~J00291$+$5934, behave in this way. The magnitudes of the phase
residuals of these \mbox{AMXPs} are consistent with the nearly aligned
moving spot model if the emitting areas wander in the azimuthal
direction by distances $\sim\,$$10^{-2}$--$10^{-3}$ times the stellar
circumference.

A second expectation in the nearly aligned moving spot model is that the
arrival times of pulses will form a track in the phase-residual versus
pulse-amplitude plane, especially if the change in the pulse amplitude
is large. Such a track will be formed if the emitting areas where the
accreting matter impacts the stellar surface move repeatedly along a
particular path in stellar latitude and longitude as the accretion rate
and the structure of the inner disk change. This is expected because
models of the flow of accreting matter from the inner disk to the
stellar surface predict that matter will be guided along particular but
different magnetic flux tubes as the accretion rate and the structure of
the inner disk vary (see Section~\ref {sec:modeling-emission}). As
discussed in Section~\ref{sec:correlated-variations}, tracks of this
type are observed in plots of pulse arrival time versus pulse amplitude
for XTE~J1814$-$338 and XTE~J1807$-$294.

\textit {Accretion- and nuclear-powered oscillations}. The success of
the emitting spot model discussed here supports magnetic field evolution
models in which the magnetic flux of the accreting neutron star becomes
concentrated near its spin axis as it is spun up. As discussed in
Sections~\ref{sec:accretion-nuclear} and~\ref{sec:pole-movement}, these
evolutionary models can produce magnetic fields strong enough to
partially confine accreting nuclear fuel near the star's magnetic poles,
even though the dipole component of the magnetic field is very weak.
This picture of magnetic field evolution in turn suggests that the
shapes and phases of the nuclear- and accretion-powered pulses are
similar to one another in some \mbox{AMXPs} because the nuclear- and
accretion powered X-ray emission comes from approximately the same area
on the stellar surface.

\textit {Effects of rapid spot motions}. In
Section~\ref{sec:spot-movements}, we discussed the effects of spot
movements on timescales shorter than the time required to construct a
pulse profile. Such effects are expected, because emitting spots are
likely to move on the stellar surface on timescales at least as short as
the $\sim\,$0.1~ms dynamical time there whereas constructing a pulse
profile usually requires folding $\sim\,$$10^{2}$--$10^{3}$~s of X-ray
flux data. Rapid spot motions will produce X-ray flux variations on
these same timescales. Our computations show that motion of the emitting
area on the stellar surface on timescales longer than the spin period
usually changes the amplitudes and the phases of the harmonic components
of the theoretical pulse profile.

Variations of the X-ray flux on any timescales shorter than the time
required to construct a pulse profile will appear in the analysis as
noise in excess of the normal counting noise. This noise will reduce the
measured amplitude of the oscillations at the spin frequency and its
overtones for all spot locations, but its effect is likely to be
stronger when the emitting area is near the spin axis because
displacement of the emitting area by a given distance there produces a
larger change in the pulse phase and, for many geometries, in the pulse
amplitude. This effect will therefore tend to reduce the apparent pulse
amplitude even further when the emitting area is close to the spin axis.

\textit {Undetectable and intermittent pulsations}. In Section~\ref
{sec:spot-inclination}, we showed that if the emitting areas of some
\mbox{AMXPs} are very close to the spin axis and remain there, the
amplitudes of the oscillations they would produce can be $\sim\,$0.5\%
or less, making them undetectable with current instruments. Rapid X-ray
flux variations will make accretion-powered oscillations at the spin
frequency more difficult to detect. Other effects, such as scattering of
X-ray photons in circumstellar gas, may also play a role in reducing the
detectability of such oscillations. These results show that the nearly
aligned moving spot model may, possibly in combination with other
effects, explain the nondetection of accretion-powered oscillations at
the millisecond spin frequencies of some accreting neutron stars in
which nuclear-powered oscillations have been detected. In Section~\ref
{sec:amplitude-variations}, we showed that if the emitting area is
within a few degrees of the spin axis and moves toward the rotation
equator by $\sim\,$10$\arcdeg$, oscillations that were undetectable can
become detectable. This may explain why accretion-powered oscillations
appear only intermittently in some \mbox{AMXPs} \citep{lamb08a}. The
model suggests that oscillations may also disappear intermittently in
some \mbox{AMXPs}.

\textit {Evolution of \mbox{AMXP} magnetic fields}. In
Section~\ref{sec:pole-movement}, we explained why the magnetic poles of
most \mbox{AMXPs} are expected to be very close to their spin axes. One
consequence is that their magnetic fields will channel accreting gas to
the stellar surface near the spin axis. Hence the star's
accretion-powered X-ray emission will come from areas near the spin
poles. A second consequence is that many \mbox{AMXPs} may have surface
magnetic fields as strong as $\sim\,$$10^{11}$--$10^{12}$~G, even though
the dipole components of these fields are only
$\sim\,$$10^{8}$--$10^{9}$~G. If so, many \mbox{MRPs} may also have
surface magnetic fields as strong as $\sim\,$$10^{11}$--$10^{12}$~G,
even though the dipole components inferred from their spin-down rates
are only $\sim\,$$10^{8}$--$10^{9}$~G.

\textit {Transient nature of \mbox{AMXPs}}. In
Section~\ref{sec:transients}, we noted that the picture of \mbox{AMXP}
magnetic field evolution just described suggests why the \mbox{AMXPs} in
which accretion-powered oscillations have been detected are in transient
systems. If the magnetic poles of most neutron stars in \mbox{LMXBs}
were forced very close to their spin axes during the initial, persistent
phase of mass transfer, accretion-powered X-ray oscillations would be
difficult or impossible to detect. Later, when mass transfer is
transient, the stars will spin down, causing their magnetic poles to
move away from their spin axes and making accretion-powered X-ray
oscillations detectable.

\textit {Tests of the model}. The nearly aligned moving spot model leads
to a number of expectations about \mbox{AMXP} X-ray emission that can be
tested (see Section~\ref{sec:discussion} for details):

\begin{enumerate}
\item  The amplitudes, harmonic content, and arrival times of pulses
should be functions of the X-ray luminosity and spectrum of the pulsar.

\item  The amplitudes, harmonic content, and arrival times of pulses
should be correlated with one another.

\item  The arrival times of pulses with low amplitudes should fluctuate
much more than the arrival times of pulses with high amplitudes.

\item  Pulse phase residuals are likely to form a track in the
phase-residual versus pulse-amplitude plane. Such tracks are likely to
be more evident if the range of the pulse amplitude variation is large.
The position of pulses along such a track should be correlated with the
X-ray luminosity and spectrum of the pulsar.

\item  If the accretion- and nuclear-powered pulses of an \mbox{AMXP}
appear nearly identical, long-term (days-to-weeks) variations in the
phase residuals of the two types of pulses should track one another.

\item  The strength of the excess background noise produced by pulse
shape fluctuations should be correlated with the amplitudes, harmonic
content, and arrival times of pulses and the X-ray luminosity and
spectrum of the pulsar.

\item  The excess noise produced by motion of the emitting area should
be stronger when the pulse amplitude is smaller and weaker when the
pulse amplitude is larger.

\item  If \mbox{AMXPs} do have surface magnetic fields as strong as
$\sim\,$$10^{11}$--$10^{12}$~G, their keV X-ray spectra may show
strong-magnetic-field features. Such features are more likely in
\mbox{AMXPs} that show evidence of nuclear fuel confinement, such as
SAX~J1808.4$-$3658 and XTE~J1814$-$338.

\item  If \mbox{AMXPs} have surface magnetic fields
$\sim\,$$10^{11}$--$10^{12}$~G, their offspring should have millisecond
spin periods and total magnetic fields of similar strength, but dipole
fields $\sim\,$$10^{8}$--$10^{9}$~G. This should affect particle
acceleration and $\gamma$-ray emission by these neutron stars.

\item  If the explanation of the transient nature of the \mbox{AMXPs}
suggested here is correct, they should be spinning down on long
timescales.
\end{enumerate}

\textit{Note added in manuscript:} This work was presented at the 2008
April Amsterdam Workshop on Accreting Millisecond Pulsars, where we
emphasized a test of our nearly aligned moving spot model that could
also have wider implications. We argued that the close similarity of the
accretion-powered and burst oscillation waveforms in XTE~J1814$-$338
strongly suggests that the emitting regions that produce them are
similar and collocated, and that if, as we had proposed previously, the
phase wandering of the accretion-powered oscillations is caused by
movement of the emitting region, then the phase of the burst
oscillations should wander in the same way. After the Workshop,
\citet{watt08} investigated this possibility and found just such a
correlation in the \textit {RXTE} data on XTE~J1814$-$338. In
particular, these authors found that during the main part of the
two-month outburst of XTE~J1814$-$338, its burst oscillation not only
has a waveform similar to that of the accretion-powered oscillation
\citep{stro03, watt05, watt06}, but is also phase-locked with it, and
that the peak of the burst oscillation coincides with the peak of the
soft component of the accretion-powered oscillation. This indicates
that, as we had suggested, the accretion- and nuclear-powered emitting
regions in this pulsar very nearly coincide, and that the simultaneous
wandering of the arrival times of both oscillations by $\sim\,$1~ms
($\sim\,$0.3 in phase) during the outburst is due to wandering of the
matter (and hence the fuel) deposition pattern on the stellar surface.

\acknowledgments
We thank Deepto Chakrabarty, Paul Demorest, Jacob Hartman, Mike Muno,
Alessandro Patruno, Scott Ransom, Ingrid Stairs, Michiel van der Klis,
and Anna Watts for helpful discussions. We thank Anthony Chan and Aaron
Hanks for assistance in analyzing and plotting our results. The results
presented here are based on research supported by NASA grant
NAG~5-12030, NSF grant AST0709015, and funds of the Fortner Endowed
Chair at Illinois, and by NSF grant AST0708424 at Maryland.

\clearpage

\end{document}